\documentclass[12pt]{article}

\usepackage{slashed}
\usepackage{color, verbatim}
\usepackage{latexsym}
\usepackage{amsmath,accents}

\usepackage{cite}
\usepackage{hyperref}

\usepackage{amssymb}
\usepackage{graphicx}
\usepackage{bm}
\usepackage{hyperref}
\usepackage{adjustbox}

\makeatletter

\@addtoreset{equation}{section}
\makeatother

\newcommand{\be}{\begin{equation}}
\newcommand{\ee}{\end{equation}}
\newcommand{\bea}{\begin{eqnarray}}
\newcommand{\eea}{\end{eqnarray}}

\begin{document}

\thispagestyle{empty}

\hfill FTUV-17-0222.4164

\hfill IFIC/17-08

\begin{center}

\begin{center}

\vspace{1.5cm}

{\Large\sc A light sneutrino rescues the light stop}\\

\end{center}

\vspace{0.8cm}

\textbf{
  M. Chala$^{\,a}$, A. Delgado$^{\,b}$,  G. Nardini$^{\,c}$, M. Quir\'os$^{\,d}$}\\

\vspace{1.cm}
${}^a\!\!$ {\em {Departament de F\'isica T\`eorica, Universitat de Val\`encia and IFIC, Universitat de Val\`encia-CSIC, Dr. Moliner 50, E-46100 Burjassot (Val\`encia), Spain}}

\vspace{.1cm}
${}^b\!\!$ {\em {Department of Physics, University of Notre Dame\\Notre Dame, IN 46556, USA}}

\vspace{.1cm}
${}^c\!\!$ {\em {Albert Einstein Center (AEC), Institute for Theoretical Physics (ITP),\\ University of Bern, Sidlerstrasse 5, CH-3012 Bern, Switzerland}}

\vspace{.1cm}
${}^d\!\!$ {\em {Institut de F\'{\i}sica d'Altes Energies (IFAE),\\ The Barcelona Institute of  Science and Technology (BIST),\\
Instituci\'o Catalana de Recerca i Estudis  
Avan\c{c}ats (ICREA)\\ Campus UAB, 08193 Bellaterra (Barcelona) Spain}}

\end{center}

\centerline{\bf Abstract}
\vspace{2 mm}
\begin{quote}\small
  Stop searches in supersymmetric frameworks with $R$-parity
  conservation usually assume the lightest neutralino to be the
  lightest supersymmetric particle. In this paper we consider an
  alternative scenario in which the left-handed tau sneutrino is
  lighter than neutralinos and stable at collider scales, but possibly
  unstable at cosmological scales. Moreover the (mostly
  right-handed) stop $\widetilde t$ is lighter than all
  electroweakinos, and heavier than the scalars of the third
  generation doublet, whose charged component, $\widetilde\tau$, is heavier
  than the neutral one, $\widetilde\nu$. The remaining supersymmetric
  particles are decoupled from the stop phenomenology. In most of the
  parameter space, the relevant stop decays are only into $t
  \widetilde\tau \tau$, $t\widetilde\nu\nu$ and $b \widetilde\nu \tau$
  via off-shell electroweakinos. We constrain the branching ratios of
  these decays by recasting the most sensitive stop searches. Due to
  the ``double invisible'' kinematics of the $\widetilde t\to
  t\widetilde\nu\nu$ process, and the low efficiency in tagging the
  $t\widetilde\tau\tau$ decay products, light stops are generically
  allowed. In the minimal supersymmetric standard model with $\sim$
  100~GeV sneutrinos, stops with masses as small as $\sim$ 350\,GeV
  turn out to be allowed at 95\% CL.

\end{quote}

\vfill

\section{Introduction}
\label{sec:introduction}

In most supersymmetric (SUSY) models, $R$-parity conservation is
implemented to avoid rapid proton decay, which implies that the
lightest supersymmetric particle (LSP) is stable. As there are strong
collider and cosmological constraints on long-lived charged
particles~\cite{Barate:1997dr,Abreu:2000tn,Abbiendi:2003yd,Achard:2001qw,Kopeliansky:2015gbi,Khachatryan:2016sfv},
the LSP is preferably electrically neutral. This, together with the
appealing cosmological features of the neutralino, has had a strong
influence on the ATLAS and CMS choice on the SUSY searches. Most of
them indeed assume the lightest neutralino to be the LSP or,
equivalently for the interpretation of the LHC searches, the
long-lived particle towards which all produced SUSY particles decay
fast.

Searches under these assumptions are revealing no signal of new
physics and putting strong limits on SUSY models.  The interpretation
of these findings in simplified models provides lower bounds at around
900 and 1800\,GeV for the stop and gluino masses,
respectively~\cite{Bianchi:2242387, Kazana:2016gni}, which are in
tension with naturalness in supersymmetry. In this sense, the bias for
the neutralino as the LSP, as well as an uncritical understanding of
the simplified-model interpretations, is driving the community to believe that supersymmetry
can not be a natural solution to the hierarchy problem anymore. In the
present paper we break with this attitude and take an alternative
direction: we assume that \emph{the LSP is not the lightest neutralino
  but the tau sneutrino}~\footnote{For further studies along similar
    directions, see e.g.~Refs.~\cite{Cerdeno:2008ep,Arina:2015uea,
    Arina:2013zca}.}. Moreover we avoid peculiar simplified model
assumptions and deal with realistic, and somewhat non trivial,
phenomenological scenarios. As we will see, the findings in this
alternative SUSY scenario make it manifest the strong impact that
biases have on our understanding on the experimental bounds and, in
turn, on the viability of naturalness.
      
As the lightest neutralino is not the LSP, we focus on scenarios with all 
gauginos (gluinos and electroweakinos) heavier than some scalars. These
scenarios, discussed in the context of natural supersymmetry, are
feasible in top-down approaches, as e.g.~in the following supersymmetry
breaking mechanisms.
   \begin{description}
   \item[Gauge mediation] \hfill \break In gauge mediated
     supersymmetry breaking (GMSB)~\cite{Giudice:1998bp} the ratio of
     the gaugino ($m_{1/2}$) over the scalar ($m_0$) masses behaves
     parametrically as $m_{1/2}^2/m_0^2\propto N f(F/M^2)$, where $N$
     is the number of messengers, $F$ the supersymmetry breaking
     parameter and $M$ the messenger mass.  The condition
     $F/M^2\lesssim 1$ guarantees the absence of tachyons in the
     messenger spectrum, and if saturated, it yields $f\simeq 3$. In
     this way, for large $N$ or $F/M^2$ close to one, the hierarchy
     $m_{1/2}\gg m_0$ emerges.  Within this hierarchy, gluinos are
     heavier than electroweakinos, and stops heavier than staus,
     parametrically by factors of the order of $g_s^2/g_\alpha^2$ at
     the messenger mass scale $M$, with $g_\alpha$ being the relevant gauge coupling. 
          The
     renormalization group running to low scales increases these mass
     splittings for $M$ much above the electroweak scale.  Further
     enhancements to these mass gaps can be achieved by including also
     gravity mediation contributions or extending the standard model
     (SM) group under which the messengers
     transform~\cite{Delgado:2012rk}~\footnote{In particular, we
       assume that the slepton singlet $\widetilde\tau_R$ is much
       heavier than the slepton doublet $(\widetilde{\nu}, \widetilde
       \tau)_L$. In GMSB scenarios this hypothesis can be fulfilled
       only if the messengers transform under a
       beyond-the-standard-model group with e.g.~an extra $U(1)$ such
       that the extra hypercharge of the lepton singlet is, in
       absolute value, larger than the one of the lepton doublet. For
       instance if we extend the SM gauge group by a $\widetilde
       U(1)$, with hypercharge $\widetilde Y$, from $E_6$ one can
       easily impose the condition that $\widetilde Y(\nu_L)=0$ while
       $\widetilde Y(\tau_R)\neq
       0$~\cite{delAguila:1986klm,Langacker:2008yv}. In this model one
       needs to enlarge the third generation into the $27$ fundamental
       representation of $E_6$ decomposed as $27=16+10+1$ under
       $SO(10)$, while $16=10+\bar 5+\nu^c$ and $10=5_H+\bar 5_H$
       under $SU(5)$. Then we get $4\widetilde Y=(-1,0,-2,2,1,-3)$ for
       the $SU(5)$ representations $(10,\bar 5,\nu^c,5_H,\bar 5_H,1)$,
       respectively.}.

   \item[Scherk-Schwarz] \hfill \break In five-dimensional SUSY
     theories, supersymmetry can be broken by the Scherk-Schwarz (SS)
     mechanism~\cite{Scherk:1979zr,Antoniadis:1990ew,Pomarol:1998sd,Antoniadis:1998sd,Delgado:1998qr,Quiros:2003gg,Dimopoulos:2014aua,Garcia:2015sfa,Delgado:2016vib}. In
     this class of theories, one can assume the hypermultiplets of the
     right handed (RH) stop and the left handed (LH) third generation
     lepton doublet localized at the brane, and the remaining ones
     propagating in the bulk of the extra dimension. In such an
     embedding, gauginos and Higgsinos feel supersymmetry breaking at
     tree level while scalars feel it through one-loop radiative
     corrections. As a consequence, the ratio between the gaugino and
     scalar masses is $m_{1/2}^2/m_{0}^2 \propto 4\pi/g_\alpha^2$. Eventually,
     gluinos and electroweakinos are very massive and almost
     degenerate, while the RH stops are light but heavier than the LH
     staus and the tauonic sneutrinos by around a factor $g_s^2/g_\alpha^2$. 
\end{description}
   
Although the aforementioned ultraviolet embeddings strengthen the
motivation of our analysis, in the present paper we do not restrict
ourselves to any particular mechanism of supersymmetry breaking. Instead we
take a (agnostic) bottom-up approach. We consider a low-energy SUSY
theory where the stop phenomenology is essentially the one of the
minimal supersymmetric standard model (MSSM) with the lighter stop
less massive than the electroweakinos and more massive than the
third-family slepton doublet~\footnote{Notice that the mass and
  quartic coupling of the Higgs do not play a key role in the stop
  phenomenology. Then, the analysis of the present paper also applies
  to extensions of the MSSM where the radiative correlation between the
  Higgs mass and stop spectrum is relaxed.}. Gluinos and the remaining
SUSY particles are heavy enough to decouple from the
collider phenomenology of the lighter stop. In this scenario the LSP at collider scales
is therefore the LH \textit{tau sneutrino}. Of course, subsets of the
parameter regions we study can be easily accommodated in any of the
previously discussed supersymmetry breaking mechanisms or minor modifications
thereof.
    
\begin{figure}[htb]
\begin{center}
  \includegraphics[height=0.25\columnwidth]{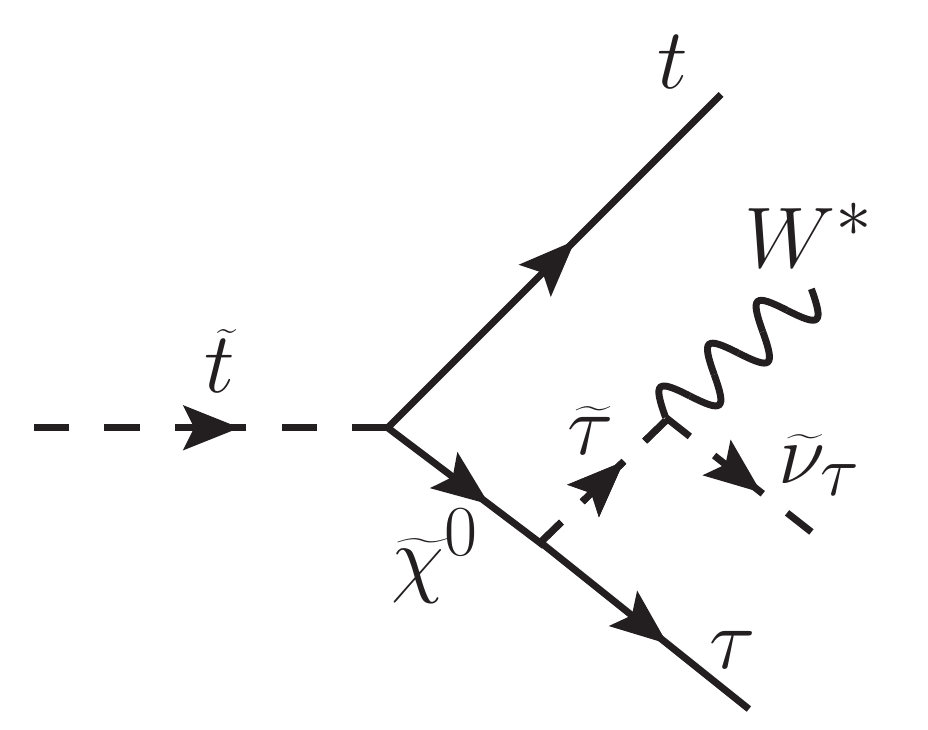}
\includegraphics[height=0.25\columnwidth]{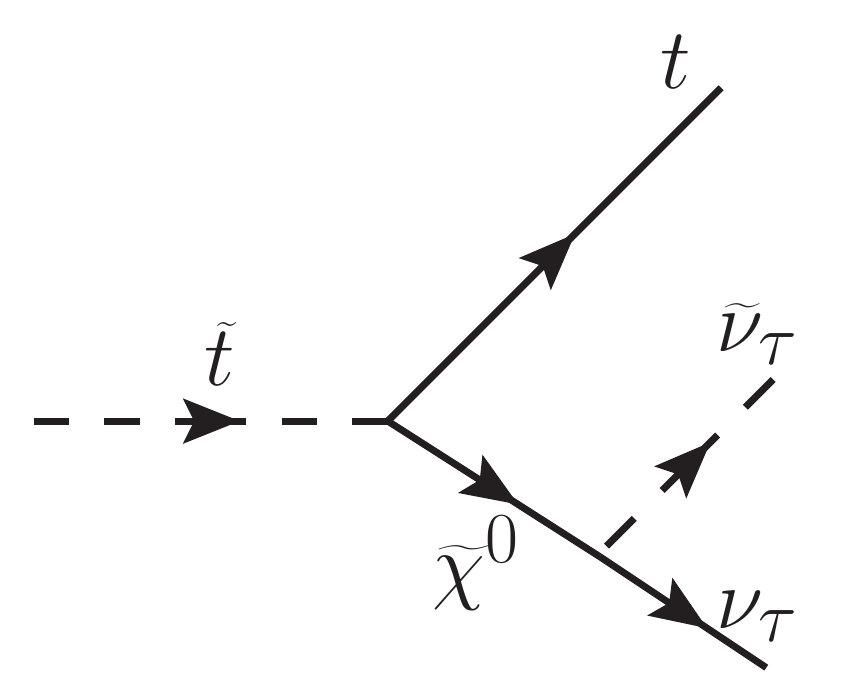}
\includegraphics[height=0.25\columnwidth]{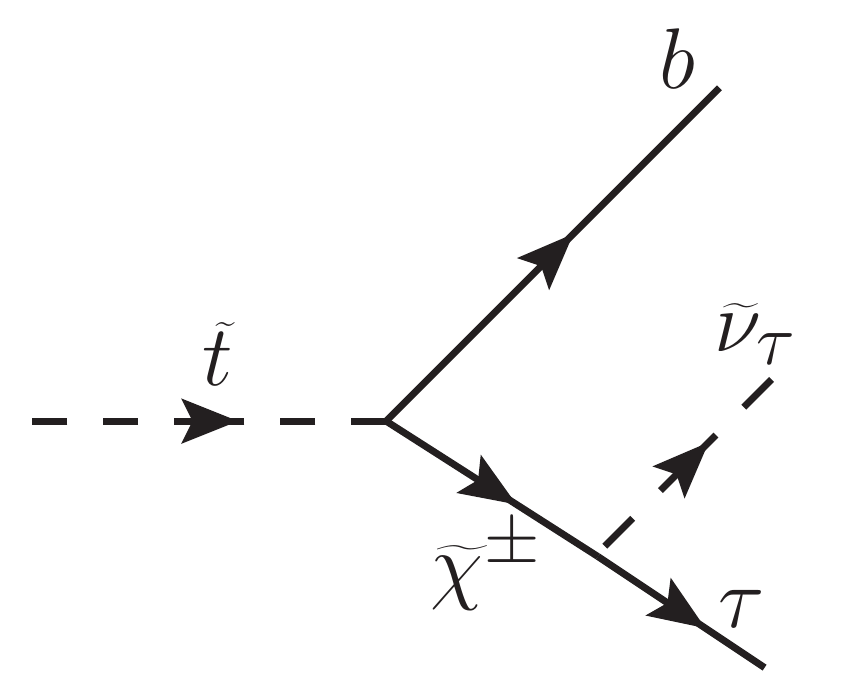}
\end{center}
\caption{\it The dominant stop decays in our analysis. Left diagram:
  Production of a top, a tau and a stau, promptly decaying into soft
  $W$-boson products and missing transverse energy. Middle diagram:
  Production of a top and two correlated sources of missing transverse
  energy. Right diagram: Production of a bottom, tau and missing
  transverse energy.}\label{fig:decays}
\end{figure}

In the considered parameter regime, the phenomenology of the lighter
stop, $\widetilde t$, is dominated by three-body decays via off-shell
electroweakinos into staus and tau neutrinos, $\widetilde\tau$ and
$\widetilde\nu$. The viable decay channels are very limited. If the
masses of the lightest sneutrino and the lighter stop are not
compressed, the only potentially relevant stop decays are $\widetilde
t \to t \widetilde\nu \nu$, $\widetilde t \to t \widetilde\tau \tau$,
$\widetilde t \to b \widetilde\nu \tau$ and $\widetilde t \to b
\widetilde\tau \nu$, the latter being negligible when the interaction
between the lighter stop and the Wino is tiny (see more details in
Sec.~\ref{sec:model})~\footnote{As a practical notation, we are not
  differentiating particles from antiparticles when indicating the
  decay final states.}. Thus, for scenarios where the lighter stop has
a negligible LH component and/or the Wino is close to decoupling, the
relevant stop signatures reduce to those depicted in
Fig.~\ref{fig:decays}. This is the stop phenomenology we will
investigate in this paper.

A comment about dark matter (DM) is here warranted. It is well known
that the LH sneutrino is not a good candidate for \emph{thermal} DM~\cite{Falk:1994es,Arina:2007tm}, as it is ruled out by direct detection
experiments~\cite{Tan:2016zwf,Akerib:2016vxi}. Therefore, in a model like the one we
study here, one needs a different approach to solve the DM
problem. Since many of the available approaches would modify the
phenomenology of our scenario only at scales irrelevant for collider
observables, incorporating such changes would not modify our results
(for more details see Sec.~\ref{sec:conclusions}).

The outline of the paper is the following. In Sec.~\ref{sec:model} we
provide further information on the scenario we consider, and on the
effects that the electroweakino parameters have on the stop
signatures. In Sec.~\ref{sec:analyses} we single out the ATLAS and CMS
analyses that, although performed to test different frameworks, do
bound our scenario. The consequent constraints on the stop branching
ratios and on stop and sneutrino masses are presented in the same
section. The implications for some benchmark points and the viability
of stops as light as 350\,GeV are explained in
Sec.~\ref{sec:constraints}. Sec.~\ref{sec:conclusions} reports on the
conclusions of our study, while App.~\ref{appendixA} contains the
technical details about our analysis validations.

\section{The model and dominant stop decays}
\label{sec:model}

In the MSSM and its minimal extensions, it is often considered that
naturalness requires light Higgsinos and stops, and not very heavy
gluinos. In fact, in most of the ultraviolet MSSM embeddings, the
Higgsino mass parameter, $\mu$, enters the electroweak breaking
conditions \emph{at tree level}, and only if $\mu$ is of the order of
the $Z$ boson mass the electroweak scale is naturally reproduced. This
however solves the issue only at tree level, as also the stops can
\emph{radiatively} destabilize the electroweak breaking
conditions. For this reason stops must be light, and the argument is
extended to gluinos since, when they are very heavy, they efficiently
renormalize the stop mass towards high values. Therefore stops cannot
be light in the presence of very massive gluinos without introducing
some fine tuning.

Remarkably, the above argument in favor of light Higgsinos, light
stops and not very heavy gluinos, is not general. There exist counter
examples where the Higgs sector, and thus its minimization conditions,
is independent of
$\mu$~\cite{Dimopoulos:2014aua,Garcia:2015sfa,Delgado:2016vib}, and
where heavy gluinos do not imply heavy
stops~\cite{Antoniadis:1998sd,Delgado:2016vib,Fox:2002bu}.  In view of
these ``proofs of principle'', there appears to be no compelling reason why the fundamental description of nature should not consist of a SUSY scenario with light
stops and heavy gluinos and electroweakinos. It is thus surprising
that systematic analyses on the latter parameter regime have not been
performed~\footnote{For recent theoretical analyses in the case of
  light electroweakinos and their bounds see
  e.g.~\cite{Arina:2016rbb,Han:2016xet}.}.
  
The present paper aims at triggering further attention on the subject
by highlighting that the present searches poorly constrain the stop
sector of this parameter scenario. For this purpose we focus on the
LHC signatures of the lighter stop being mostly RH.  The illustrative
parameter choice we consider is the one where the stop and slepton
mixings are small, and the light third generation slepton doublet is
lighter than the lighter stop~\footnote{These features naturally
  happen in GMSB and SS frameworks. For GMSB, the trilinear parameter
  $A$ arises at two loops whereas $m_0$ appears at one loop. Thus the
  ratio $A/m_0$ is one-loop suppressed. Similarly, the SS breaking
  produces a large tree-level mass for the LH stop and the RH stau
  fields in the bulk, and generates $A$ at one loop, such that $A/m_0$
  is small due to a one-loop factor. Moreover, the ratio
  $m_{\widetilde\nu}/m_{\widetilde t}$ is parametrically
  $O(g^2/g_s^2)$ in such GMSB and SS embeddings.}.  The remaining
squarks, sleptons and Higgses are assumed to be very heavy, in
agreement with the (naive) interpretation of the present LHC
(simplified model) constraints.  Specifically, these particles, along
with gluinos, are assumed to be decoupled from the relevant light stop
phenomenology. Moreover, possible $R$-parity violating interactions
are supposed to be negligible at detector scales.

In the present parameter scenario the light stop phenomenology only
depends on the interactions among the SM particles, the lighter
(mostly RH) stop, the lighter (mostly LH) stau, the tau sneutrino and
the electroweakinos. The stop decays into sleptons via off-shell
charginos and neutralinos. In principle, due to the interaction
between the stop and the neutralinos (charginos), any up--type
(down--type) quark can accompany the light stop decay
signature. Nevertheless, in practice, flavor-violating processes arise
only for a very compressed slepton-stop mass spectrum. For our main
purpose, which is to prove that pretty light stops are allowed in the
present scenario, the analysis of this compressed region is not
essential~\footnote{Notice that in an extreme parameter regime, the
  stop is long lived and leads to stoponium, whose signatures are
  qualitatively different from those we are discussing
  here~\cite{Drees:1993uw, Martin:2008sv, Bodwin:2016whr}. Including this (small)
  parameter regime is irrelevant for our purposes, and we thus exclude
  it from our analysis.}. To safely avoid this region, we impose
$m_{\widetilde t}\gtrsim m_{\widetilde \nu}+70\,$GeV, with
$m_{\widetilde t}$ and $m_{\widetilde \nu}$ being the masses of the
lighter stop and the tau sneutrino, respectively.

The kinematic distributions associated to the stop decays strongly
depend on the stau and sneutrino masses. In particular, the sneutrino mass
$m_{\widetilde \nu}$ is free from any direct constraint coming from
collider searches and, as stressed in Sec.~\ref{sec:introduction}, we
refrain from considering bounds that depend on cosmological scale
assumptions. On the other hand, numerous collider-scale dependent
observables affect the stau as we now discuss.

The ALEPH, DELPHI, L3 and OPAL Collaborations interpreted the LEP data
in view of several SUSY scenarios and, depending on the different searches, they
obtain the stau mass bound $m_{\widetilde \tau}\gtrsim
90\,$GeV~\cite{Barate:1997dr,Abreu:2000tn,Abbiendi:2003yd,Achard:2001qw}.
A further constraint comes from the CMS and ATLAS searches for
disappearing charged tracks, for which $m_{\widetilde \tau}\simeq
90\,$GeV is ruled out if the stau life-time is
long~\cite{Kopeliansky:2015gbi,Khachatryan:2016sfv}. However, in the
present scenario with small sparticle mixings, the mass splitting
$m_{\widetilde \tau}-m_{\widetilde \nu}$, given by
\be
    m_{\widetilde\tau}^2-m_{\widetilde\nu}^2=\frac{\tan^2\beta-1}{\tan^2\beta+1}\cos^2\theta_W m_Z^2 +\mathcal O(m_\tau^2)~,
\ee
can be sufficiently large to lead to a fast stau decay, and in fact
the charged track LHC bound is eventually overcome for $m_{\widetilde
  \tau}\gtrsim 90\,$GeV and $\tan\beta >1$ (see
Sec.~\ref{sec:conclusions}).  On the other hand, a light stau with
mass close to the LEP bound modifies the 125 GeV Higgs signal strength
$\mathcal R(h\to \gamma\gamma)$ unless $\tan\beta\ll 100$
~\cite{Carena:2012gp}. All together these bounds hint at an
intermediate (not very large) choice of $\tan\beta$, as
e.g.~$\tan\beta\sim10$.

Finally, a light stau, as well as a light stop, can modify the
electroweak precision observables~\cite{Marandella:2005wc}. One
expects the corresponding corrections to be within the experimental
uncertainties for $m_{\widetilde \tau}\gtrsim 90\,$GeV, $m_{\widetilde
  t}\gtrsim 300\,$GeV and negligible sparticle mixing, since the stop
is mostly RH and the light stau is almost degenerate in mass with the
tau sneutrino. The latter degeneracy plays a fundamental role also in
the collider signature of the stau decay: due to the compressed
spectrum, the stau can only decay into a stable (at least at detector
scales) sneutrino and an off-shell $W$ boson, giving rise to soft
leptons or soft jets.

At the quantitative level, the decay processes of the stop are
described, in the electroweak basis, by the relevant interaction
Lagrangian involving the Bino, Wino, Higgsinos, tau sneutrino, the LH
and RH stops and staus ($\widetilde B,\widetilde W, \widetilde
H_{1,2},\widetilde\nu_L$, $\widetilde t_{L,R}$ and $\widetilde
\tau_{L,R}$) as well as their SM counter-partners~\footnote{We use
  two-component Weyl spinor notation for $\psi_{L,R}$, where $\psi_L$
  are undotted spinors and $\psi_R$ dotted spinors. By definition $\bar
  \psi_{R}\equiv \psi_{R}^\dagger$ are undotted spinors.}:
\begin{align}
\mathcal L_I&=-g\left( \widetilde{t}_L^* b_L \widetilde W^+  + \widetilde\tau^*_L \nu_L \widetilde W^-
+\widetilde\nu_L^* \tau_L \widetilde W^+ \right)
-{\displaystyle\frac{g}{\sqrt{2}}}\left(  \widetilde t_L^* t_L  +\widetilde\nu_L^* \nu_L -\widetilde\tau_L^* \tau_L 
\right)\widetilde W^0\nonumber\\
&-{\displaystyle\frac{g^\prime}{\sqrt{2}}}
\left({\displaystyle \frac{1}{3}}\widetilde t_L^* t_L 
+{\displaystyle\frac{4}{3}}\widetilde t_R^* t_R +\widetilde\nu_L^*\nu_L+\widetilde\tau_L^*\tau_L-2\widetilde\tau_R^*\tau_R
\right)\widetilde B\nonumber\\
&-{\displaystyle\frac{1}{2} }\left\{{\displaystyle\frac{h_t}{\sin\beta}}\widetilde t_R^* b_L \widetilde H_2^++{\displaystyle\frac{h_b}{\cos\beta}}\widetilde t_L \bar b_R \widetilde H_1^-
+{\displaystyle\frac{h_t}{\sin\beta}}
\left(\widetilde t_R^* t_L +\widetilde t_L \bar t_R  \right) \widetilde H^0_2\right.\nonumber\\
&+\left.{\displaystyle\frac{h_\tau}{\cos\beta}}
\left[
\left(\widetilde\tau_R^*\nu_L+\widetilde\nu_L \bar\tau_R  \right) \widetilde H_1^-
-(\widetilde\tau_R^* \tau_L+\widetilde\tau_L \bar\tau_R)\widetilde H_1^0
\right]\right\}
\nonumber\\
&-i{\displaystyle\frac{g}{\sqrt{2}}}\left[(\partial^\mu\widetilde\nu_L^*) W_\mu^+\widetilde\tau+(\partial^\mu\widetilde\tau^*_L) W_\mu^- \widetilde\nu_L  \right\}+h.c.
\label{lagrangiano} ~.
\end{align}
Here $h_{t,b,\tau}$ are the SM Yukawa couplings while, following the
usual MSSM notation, $\widetilde H_2$ ($\widetilde H_1)$ is the SUSY
partner of the Higgs with up-type (down-type) Yukawa interactions. The
first two lines in Eq.~(\ref{lagrangiano}) come from $D$-term
interactions, the third and fourth lines from $F$-terms Yukawa
couplings and the last line from the covariant derivative of the
corresponding fields.
  
This Lagrangian helps to pin down the Bino, Wino and Higgsino
(off-shell) roles in the stop decays. In order to understand the
magnitude of the single contributions, it is important to remind that
the stop (stau) is mostly RH (LH). Moreover, for our scenario with
electroweakino mass parameters $M_1$, $M_2$, $\mu \gg m_Z$, the Bino,
Winos and Higgsinos are almost mass eigenstates.

The Bino and the electrically-neutral components of Winos and
Higgsinos contribute to the decays $\widetilde t \to t \widetilde\tau
\tau$ and $\widetilde t \to t \widetilde\nu \nu$ (see the first two
diagrams in Fig.~\ref{fig:decays}). We expect different branching
ratios into anti-stau tau and into stau anti-tau. This is a consequence
of the fact that the decaying particle in the first diagram of
Fig.~\ref{fig:decays} is a stop and not an anti-stop. This difference
in the branching ratios can be understood from the point of view of
effective operators obtained in the limit that the neutralinos are
enough heavy that can be integrated out. We show that this is so by
considering the two (opposite) regimes where the light stop is either
mostly RH or mostly LH.

Let us first assume that in the process $\widetilde t \to t
\widetilde\tau \tau$ the decaying stop is RH, i.e.~the field
$\widetilde t_R$ in Eq.~(\ref{lagrangiano}).  If the neutralinos are
mainly gauginos $(\widetilde B,\widetilde W^0)$, as the RH stop is an $SU(2)_L$ singlet, the process
has to be mediated by the Binos. In this case the produced top will be
RH and the lowest order (dimension-five) effective operator can 
be written as $(\widetilde t_R^* \widetilde \tau_L)( t_R \bar\tau_L)$,
by which only \textit{staus and anti-taus} are produced, but not
anti-staus and taus. For diagrams mediated by Higgsinos, the produced
top will be LH and the effective operator is $(\widetilde t_R^*
\widetilde \tau_L)( t_L \bar\tau_R)$, and again the stop decay
products are staus and anti-taus. However, in the limit of heavy
electroweakino masses, the coefficient of the latter operator is
suppressed by $\mathcal O(v/\mu)$.
Now let us instead assume that the decaying stop is LH, that is,
$\widetilde t_L$ in Eq.~(\ref{lagrangiano}). In this case the effective
operators for the exchange of gauginos and Higgsinos in $\widetilde t
\to t \widetilde\tau \tau$ would be $(\widetilde t_L^*
\widetilde\tau_R )(t_L \bar\tau_R)$ and $(\widetilde t_L^*
\widetilde\tau_R )(t_R \bar \tau_L)$ respectively, implying again that
the decay products are staus and anti-taus. The contribution to the
latter effective operators is small if the RH stau is heavy (and/or
the LH component of the stop is small), as happens in the considered
model, leading again to the production of \textit{staus and anti-taus}
with either chirality.

In reality, in our scenario with mostly RH light stops, since
neutralinos are not completely decoupled, full calculations of the
stop decays exhibit also some anti-stau and tau contributions. These
proceed from dimension-six effective operators such as
e.g.~$(\widetilde t_R^* \partial_\mu
\widetilde\tau^*_R)(t_R\bar\sigma^\mu \tau_L)$, which contain an extra
suppression factor $\mathcal O(v/\mu, v/M_{1,2})$ with respect to the
leading result. We can finally say that the decay of stops is
dominated by the production of \textit{anti-taus} while the production
of taus is chirality suppressed~\footnote{The same effect arises also
  in the $\widetilde t \to t \widetilde\nu \nu$ decay (second diagram
  in Fig.~\ref{fig:decays}), but the collider signatures of these
  different products are not relevant, for neutrinos or anti-neutrinos
  are indistinguishable at colliders.}.  Although interesting, this
effect escapes from the most constraining stop searches, which do not
tag the charge of taus or other leptons  (see
Sec.~\ref{sec:analyses}). For the purposes of the detector simulations
the stop branching ratios can thus be calculated without
differentiating the processes yielding taus or anti-taus.
     
The chirality suppression is instead crucial for the three-body decays
via off-shell charginos. In principle both decays $\widetilde t \to b
\widetilde\tau \nu$ and $\widetilde t \to b \widetilde\nu \tau$ are
allowed but, due to the chirality suppression, only the latter (which
corresponds to the third diagram in Fig.~\ref{fig:decays}) can be
sizeable in our scenario. Indeed, let us consider the case where the
stop decaying into $b_L$ and an off-shell charged Higgsino is the RH
one~\footnote{As $\widetilde t_R$ is an $SU(2)_L$ singlet it cannot
  decay via a charged gaugino $\widetilde W^\pm$.}. The only
five-dimensional effective operator that can be constructed is
$(\widetilde t_R^* \widetilde\nu_L)(b_L \bar\tau_R)$ which appears
from the mixing between $\widetilde H_2^+$ and $(\widetilde
H_1^{-})^*$, after electroweak symmetry breaking, and is thus suppressed
by a factor $\mathcal O(v/\mu)$. Now instead assume that the stop is
LH.  At leading order, the decay into $b_L$ and $\widetilde W^+$ gives
rise to the operator $(\widetilde t_L^*\widetilde \tau_L^*)(b_L
\nu_L)$~\footnote{Notice that in our convention both $b_L$ and $\nu_L$
  are undotted spinors and thus $b_L \nu_L\equiv b_L^\alpha
  \widetilde\epsilon_{\alpha\beta}\nu_{L}^\beta$, with $\widetilde\epsilon_{\alpha\beta}$
  being the Levi-Civita tensor, is Lorentz invariant.}.  Moreover, the
$\widetilde t_L$ decay into $b_R$ and $(\widetilde H_1^-)^*$ can only be
generated by a dimension-six operator which is further suppressed
by the (tiny) factor $h_b h_\tau/\cos^2\beta$. Thus, in general, only
the decay $\widetilde t \to b \widetilde\nu \tau$ can be relevant in
scenarios where the light stop is practically RH (or the Wino is much
heavier than the Higgsinos), as we are considering throughout this
work. For this reason the decay $\widetilde t \to b \widetilde\tau
\nu$ is absent in Fig.~\ref{fig:decays}, that only depicts the relevant decays
in our scenario.

In the next section we will study in detail how the present LHC data
constrain scenarios with light stops predominantly decaying into $t
\widetilde\tau \tau$, $t \widetilde\nu \nu$ and $ b \widetilde\nu
\tau$, while in Sec.~\ref{sec:constraints} we will provide some parameter
regions exhibiting this feature and relaxing the bounds on light
stops.

\section{LHC searches and the dominant decays}
\label{sec:analyses}

The data collected during the LHC Run II, even at small luminosity,
have proven to be more sensitive to SUSY signals than their
counterpart at $\sqrt{s} = 8$ TeV. Among the searches with the
most constraining expected reach, we will be interested in
those for pair-produced stops in fully hadronic final states
performed by the ATLAS and CMS Collaborations, in
Refs.~\cite{atl:2016sab,cms:2016srd}, respectively, as well
as searches for pair-produced stops in a final state with tau
leptons carried out by the ATLAS Collaboration in
Ref.~\cite{atl:2016src}. However, the results provided by these
experiments can not simply be used to constrain the signal processes
under consideration.

\begin{figure}[t]
\begin{center}
 \includegraphics[width=0.5\columnwidth]{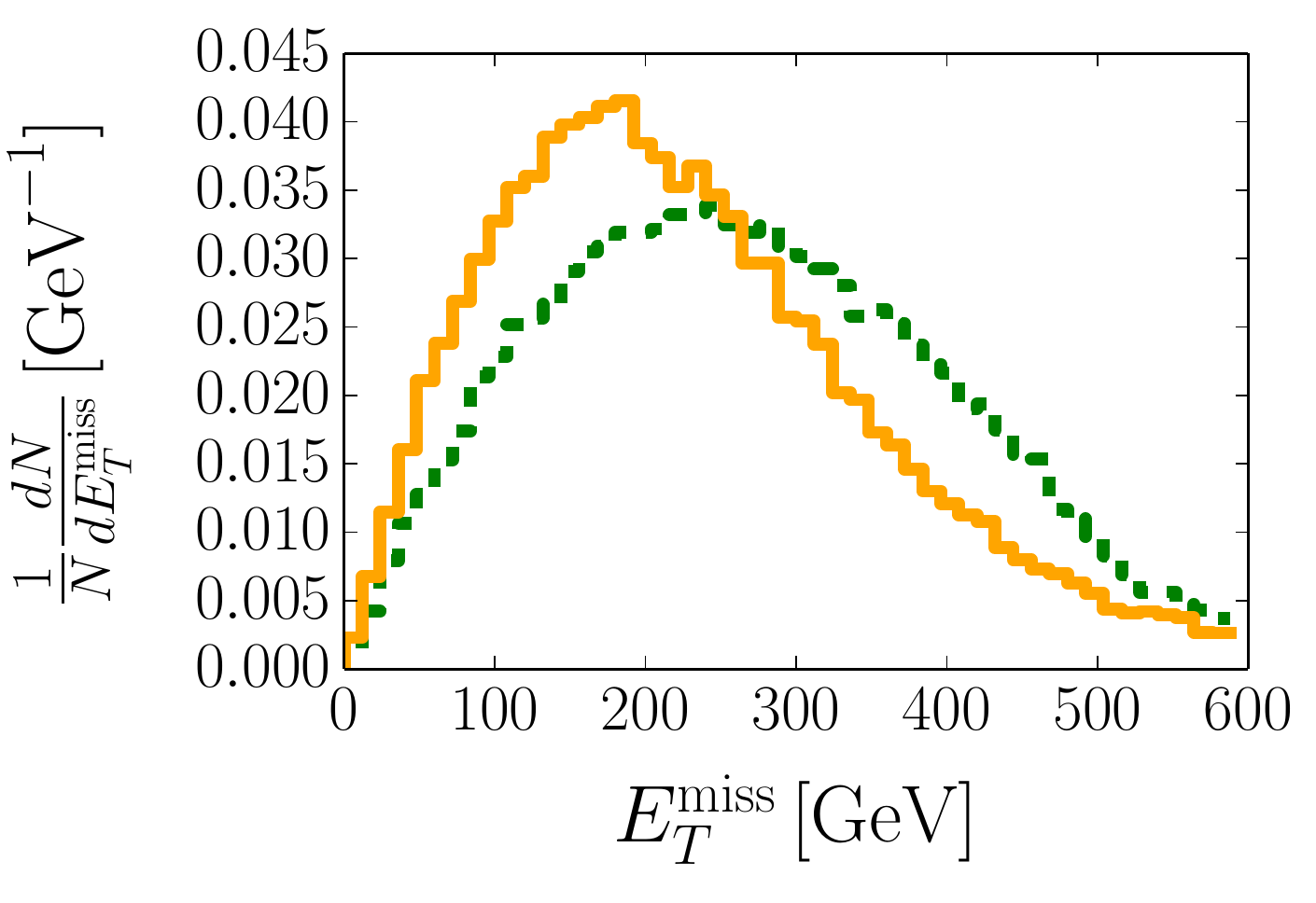}
  \includegraphics[width=0.49\columnwidth]{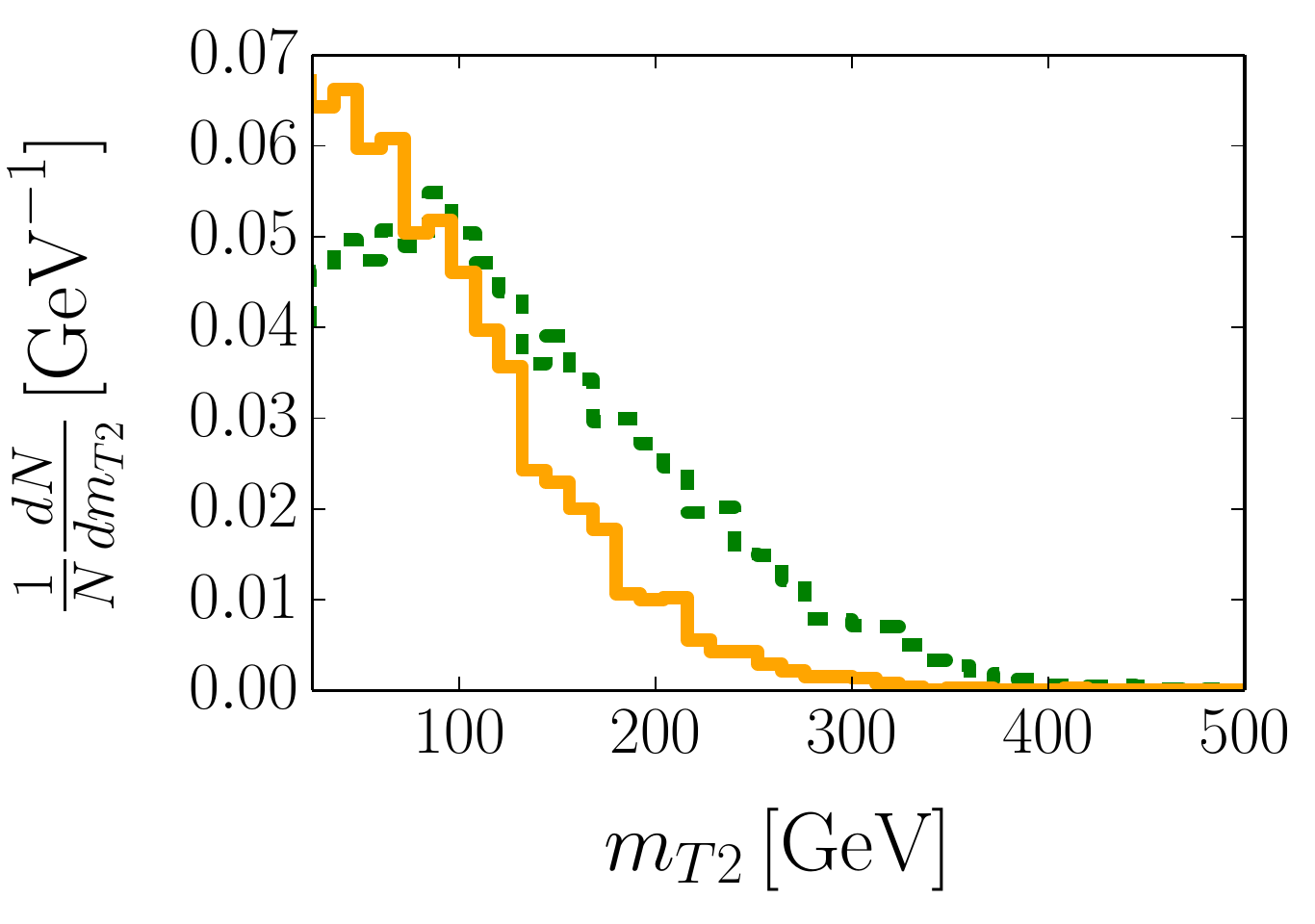}
\end{center}
\caption{\it Left panel: Normalized distribution of
  $E_{T}^{\rm{miss}}$ in the simplified model of
  Refs.~\cite{cms:2016srd, atl:2016sab} (dashed green line) and our
  scenario (solid orange line) with
  $\text{BR}(\widetilde{t}\rightarrow t \widetilde{\nu} \nu) = 1$. In
  both cases, $m_{\widetilde{t}} = 625$ GeV and $m_{\rm{LSP}} = 200$
  GeV. Right panel: Normalized distribution of $m_{T2}$ in the stop
  decay of the simplified model in Ref.~\cite{atl:2016src} (dashed
  green line) and of our scenario when
  $\text{BR}(\widetilde{t}\rightarrow b\widetilde{\nu}\tau) = 1$
  (solid orange line), with $m_{\widetilde{t}} = 700$\,GeV and
  $m_{\widetilde{\tau}} =m_{\widetilde{\nu}} = \,400$\,GeV in both
  cases.}\label{fig:distr}
\end{figure}

This reinterpretation issue is clear for the decay $\widetilde t \to t
\widetilde\tau \tau$ (see the first diagram in Fig.~\ref{fig:decays}), as
the final state is different from any other final state studied by
current searches, in particular with more taus involved. In the
$\widetilde t \to t \widetilde\nu \nu$ decay (see the
second diagram in Fig.~\ref{fig:decays}), the final state, a top plus
missing transverse energy $E_T^{\rm{miss}}$, coincides with e.g.~the
one of the $\widetilde t \to t \widetilde \chi^0$ process, with the
neutralino as the LSP studied in Refs.~\cite{atl:2016sab,cms:2016srd}. Nevertheless, since the neutralino is
\emph{off-shell} in our case, most of the discriminating variables
behave very differently, and therefore the experimental bound on
$\widetilde t \to t \widetilde \chi^0$ does not strictly
apply~\cite{Alves:2013wra}. And even the existing analyses for stops
decaying into several invisible particles, which also
Refs.~\cite{atl:2016sab,cms:2016srd,atl:2016src}
investigate, turn out to be based on kinematic cuts with
efficiencies that are unreliable in our case. This for instance holds
for the $\widetilde{t}\rightarrow b \widetilde{\nu} \tau$ decay (see the
third diagram in Fig.~\ref{fig:decays}) whose invisible particle does
not exactly mimic the ones of $\widetilde{t} \rightarrow b \tau \nu \widetilde G$
(where $\widetilde G$ is a massless gravitino) analyzed in
Ref.~\cite{atl:2016src}.

For the sake of comparison, in the left panel of Fig.~\ref{fig:distr}
we show the distributions of $E_T^{\rm{miss}}$ in the decays
$\widetilde t \to t \widetilde \chi^0$ (dashed green line) and $\widetilde t \to t
\widetilde\nu \nu$ (orange solid line) with $m_{\widetilde{t}} = 625$ GeV and
$m_{\rm{LSP}} = 200$ GeV.  In the right panel we contrast the shapes of
the transverse mass $m_{T2}$ constructed out of the tagged light
tau lepton, without any further cut, coming from the decays
$\widetilde{t}\rightarrow b \widetilde{\nu} \tau$ (dashed green line) and $\widetilde{t}
\rightarrow b \tau \nu \widetilde G$ (orange solid line) for $m_{\widetilde
  \tau}=m_{\widetilde \nu}=400\,$GeV and gravitino mass $m_{\widetilde
  G}=0$.  These kinematic variables are of fundamental importance for
the aforementioned ATLAS and CMS searches.  In particular, as
Fig.~\ref{fig:distr} illustrates, the stringent cuts on these
quantities reduce the efficiency on the signal in our model, with
respect to the standard benchmark scenarios for which the LHC searches
have been optimized. This issue was previously pointed out in
Ref.~\cite{Alves:2013wra}.

In the light of this discussion, we recast the aforementioned analyses
using homemade routines based on a combination of \texttt{MadAnalysis v5}~\cite{Conte:2012fm,Conte:2014zja} and \texttt{ROOT v5}~\cite{Brun:1997pa}, with
boosted techniques implemented via \texttt{Fastjet
  v3}~\cite{Cacciari:2011ma}.  Two signal regions, SRA and SRB, each one
containing three bins, are considered in the ATLAS fully hadronic
search~\cite{atl:2016sab} (note that SRA and SRB are not statistically
independent, though). The CMS fully hadronic
analysis~\cite{cms:2016srd} considers, instead, a signal region
consisting of 60 independent bins. Finally, the ATLAS analysis
involving tau leptons carries out a simple counting
experiment. Details on the validation of our implementation of these
three analyses can be found in App.~\ref{appendixA}. We find that our
recast of the ATLAS search for stops in the hadronic final state leads
to slightly smaller limits, while the ones of the other searches
very precisely reproduce the experimental bounds. 

\begin{table}[htb]
 \begin{center}
\begin{adjustbox}{width=0.8\textwidth}
\footnotesize
\begin{tabular}{||l|c|c|c||}\hline
 & ATLAS~\cite{atl:2016sab} (only SRB) & CMS~\cite{cms:2016srd} & ATLAS~\cite{atl:2016src}\\  
$\widetilde t \to t \widetilde\tau \tau$ &  & $\surd$ &  $\surd^*$\\ 
$\widetilde t \to t \widetilde\nu \nu$ & $\surd$ & $\surd^*$ & \\
  $\widetilde t \to b \widetilde\nu \tau$ &  & $\surd$ & $\surd^*$\\
  
\hline
 \end{tabular}
 \end{adjustbox}
 \end{center}
 \caption{\it Analyses employed for testing the different decay
   modes. The most sensitive one in each case is tagged with an asterisk.\label{tab:analyses}}
\end{table}
Thus, as shown in Tab.~\ref{tab:analyses}, we combine the whole CMS
set of bins with the above signal region SRB for probing the decay
$\widetilde t \to t\widetilde{\nu}\nu$, and with the single bin of the
ATLAS counting experiment for testing the $\widetilde t \to t
\widetilde\tau \tau$ and $\widetilde t \to b \widetilde\nu \tau$
processes~\footnote{In principle, the two ATLAS analyses could be
  combined into a single statistics. They are indeed independent, for
  one of them concentrates on the fully hadronic topology while the
  other tags light leptons. If we only combine with the CMS analysis
  is because the validation of this search gives better results. At any
  rate, no big differences are expected.}. Limits at different confidence
levels are obtained by using the CL$_s$ method~\cite{Read:2002hq}. The expected number of
background events, as well as the actual number of observed events, are
obtained from the experimental papers. Signal events, instead, result
from generating pairs of stops in the MSSM with \texttt{MadGraph
  v5}~\cite{Alwall:2014hca} that are subsequently decayed by
\texttt{Pythia v6}~\cite{Sjostrand:2006za}. The parameter cards are
produced by means of \texttt{SARAH v4}~\cite{Staub:2013tta} and
\texttt{SPheno v3}~\cite{Porod:2011nf}. When each channel is studied
separately, the corresponding branching ratio has been fixed manually to one
in the parameter card. When several channels are considered, the
amount of signal events is rescaled accordingly.

\subsection{Single channel bounds}
As discussed in the previous sections, in our scenario the possible
decay channels are $\widetilde t \to t
\widetilde\tau \tau$,
$\widetilde t \to t \widetilde\nu \nu$ and $\widetilde t \to b \widetilde\nu \tau$. In this section we consider each individual
decay channel and use the LHC data to bound the corresponding
branching ratio in the plane $(m_{\widetilde t},m_{\widetilde\nu})$.

\begin{figure}[!]
  \begin{center}
    \includegraphics[width=0.48\columnwidth]{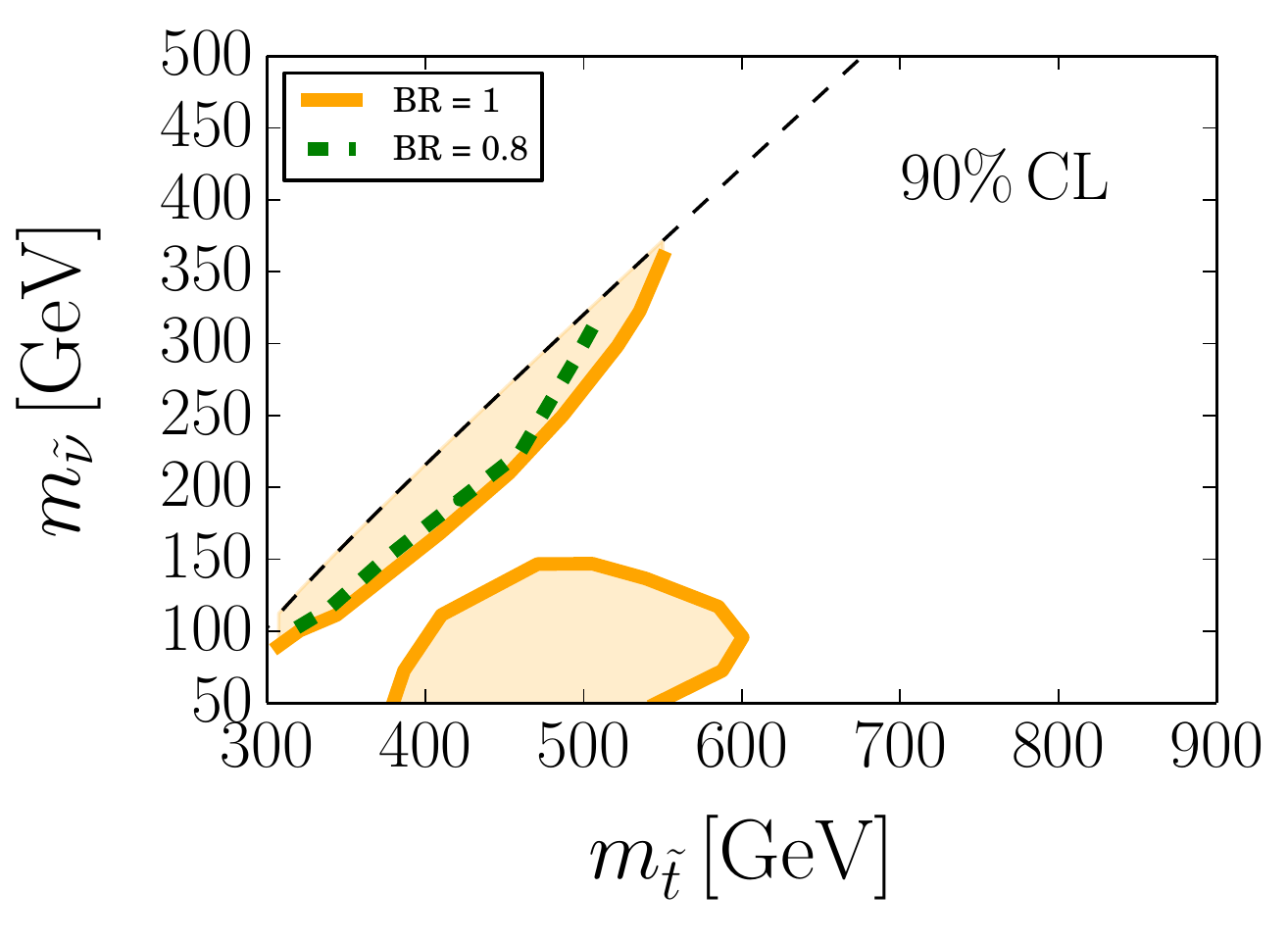}
 \includegraphics[width=0.48\columnwidth]{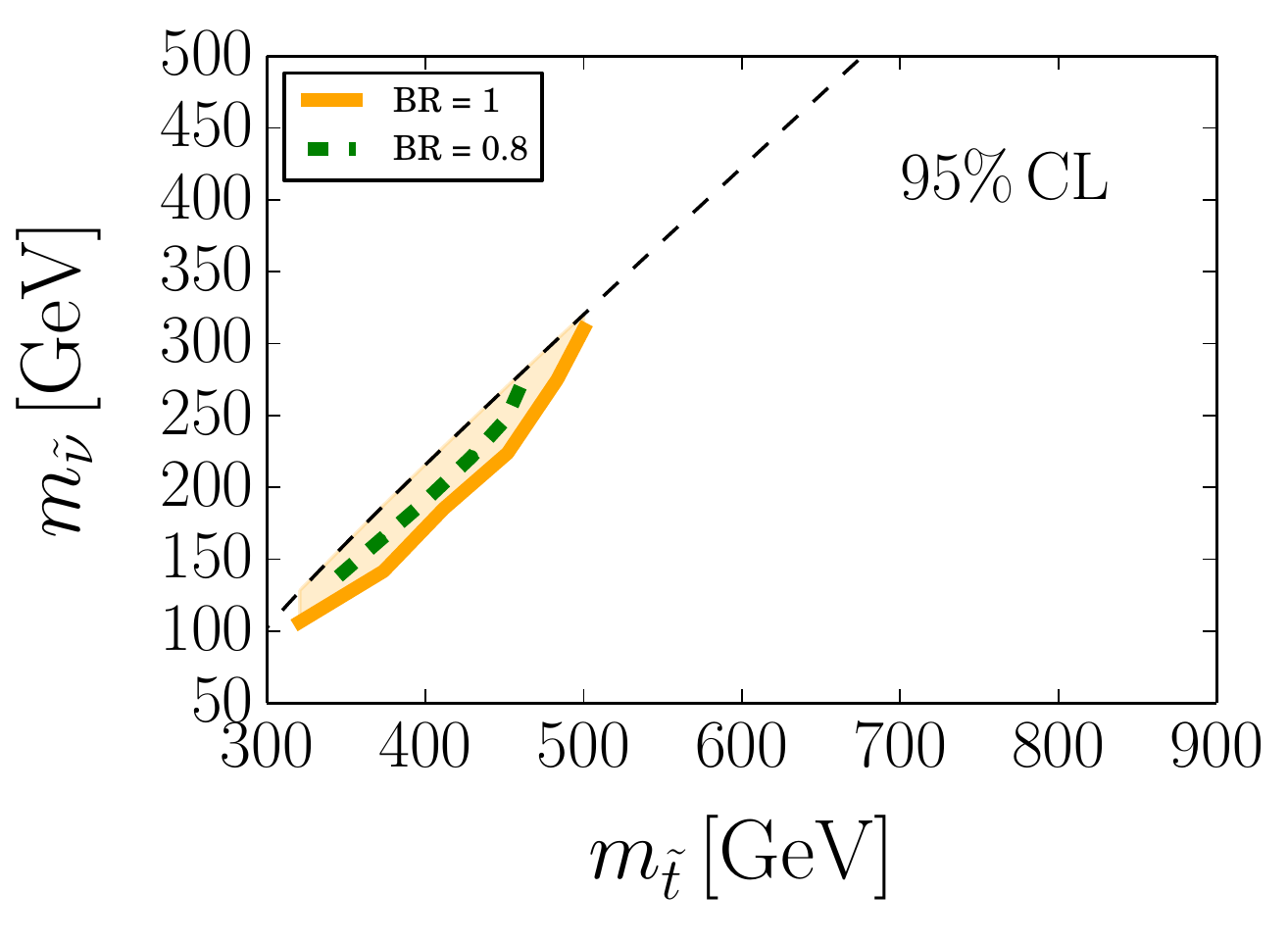}
 \includegraphics[width=0.48\columnwidth]{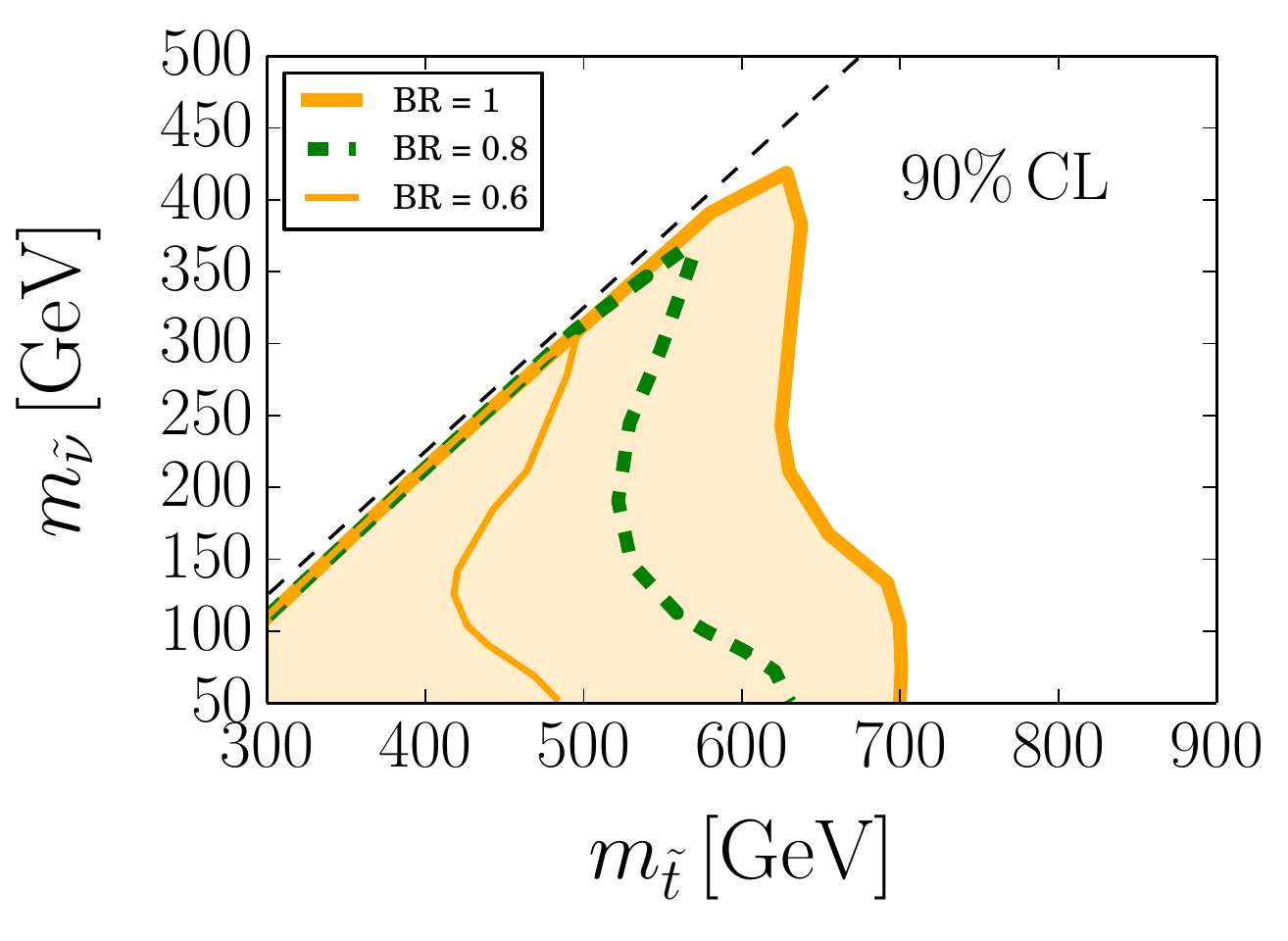}
 \includegraphics[width=0.48\columnwidth]{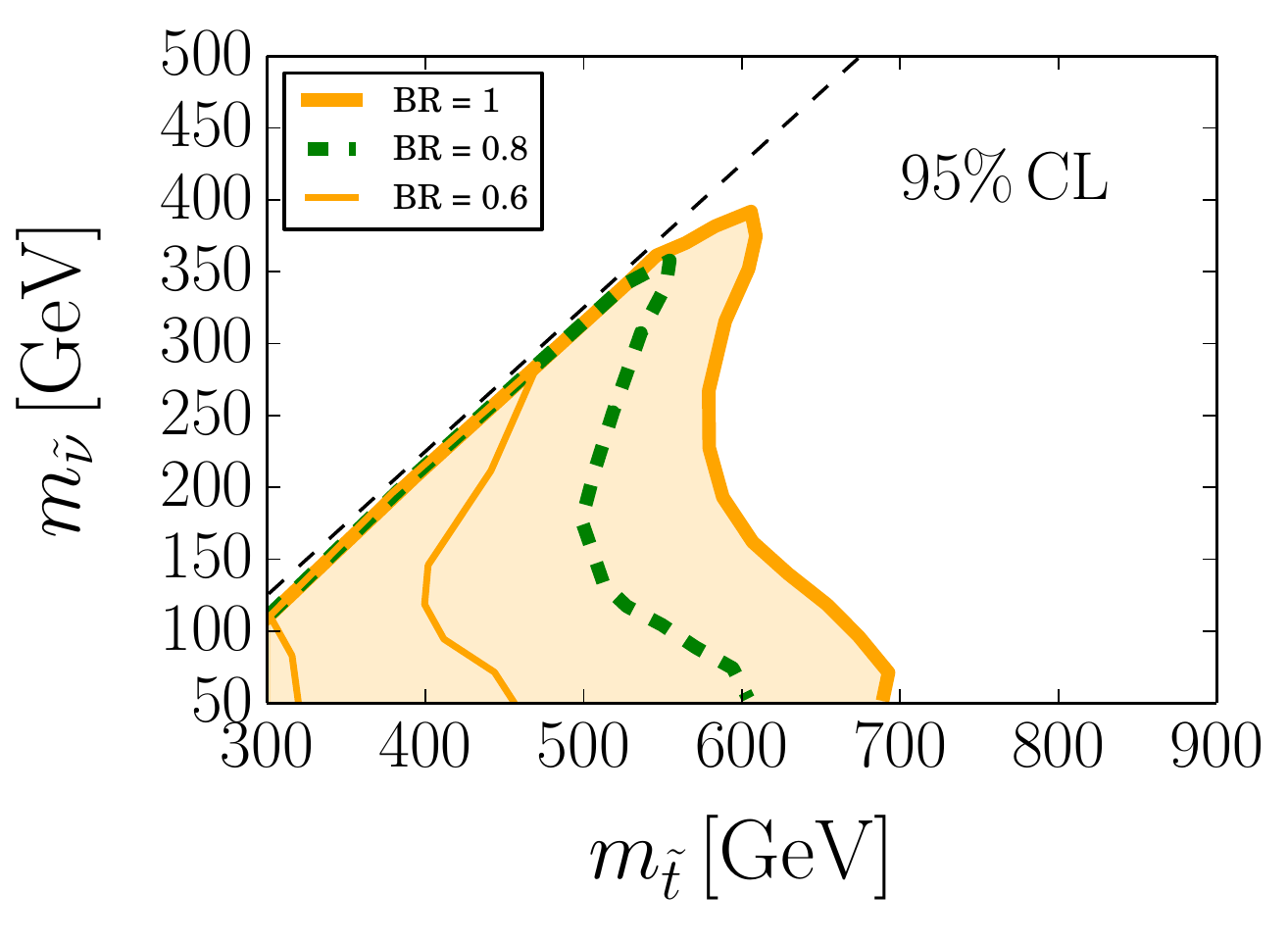}
  \includegraphics[width=0.48\columnwidth]{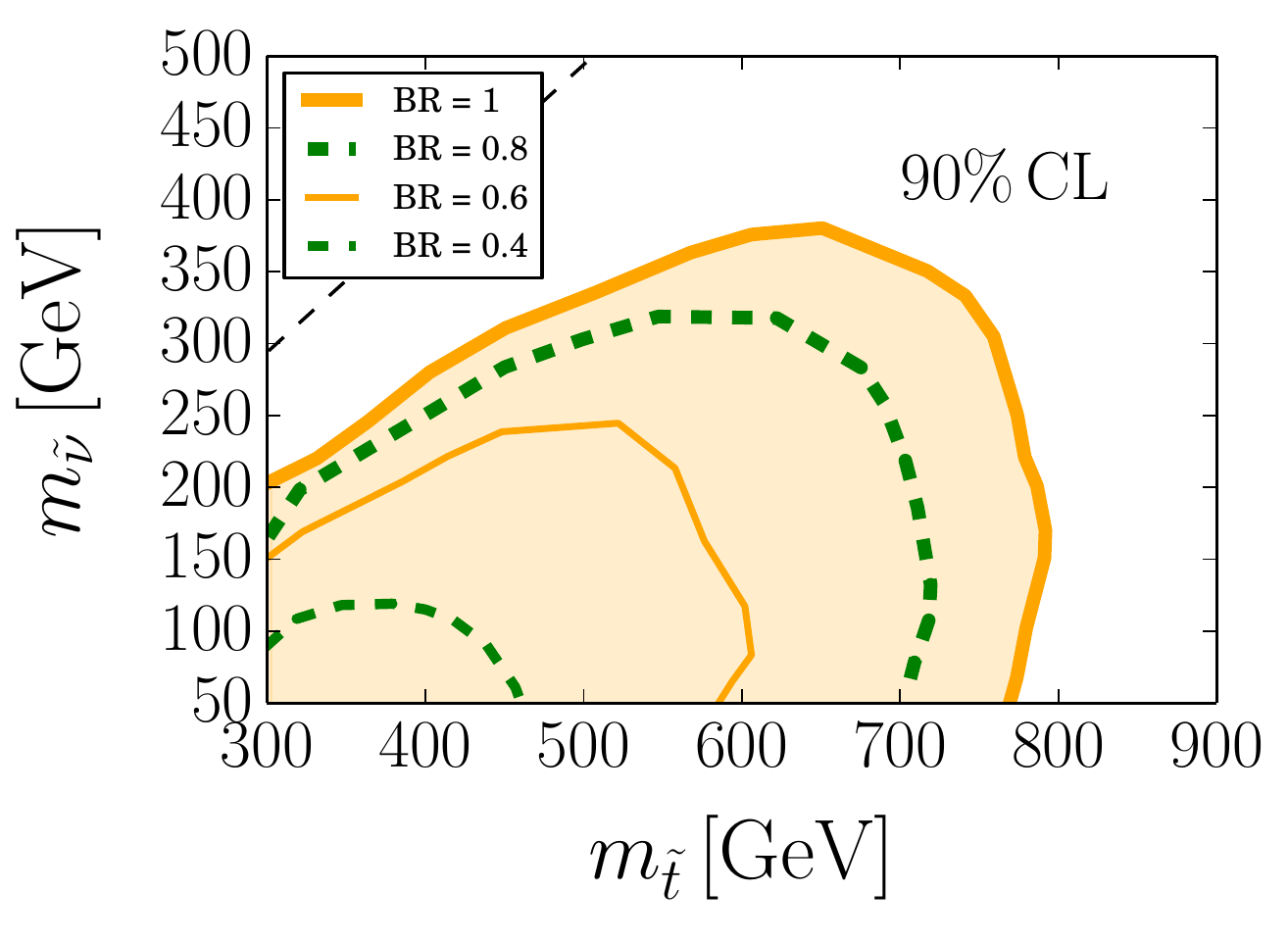}
 \includegraphics[width=0.48\columnwidth]{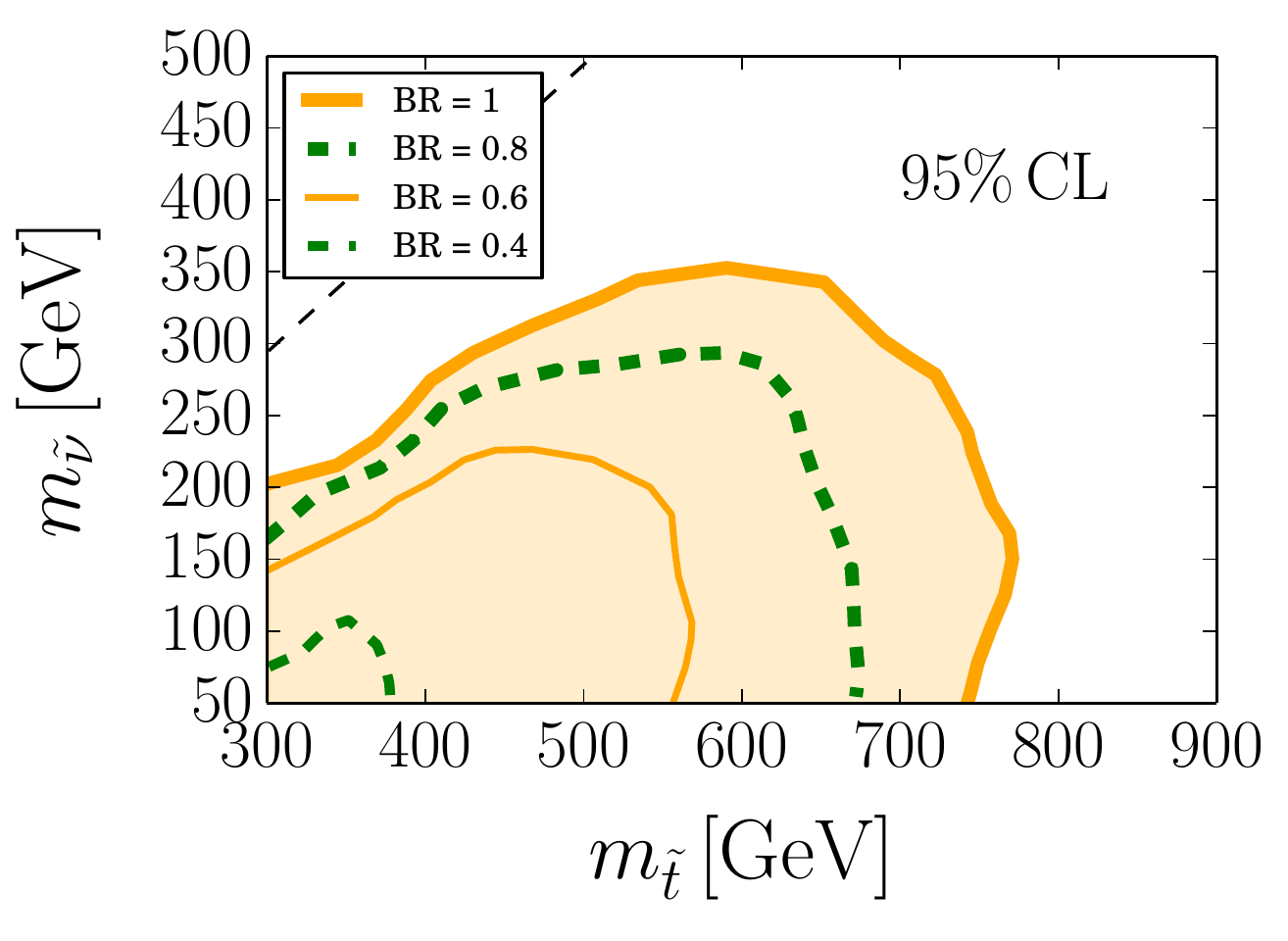}
\end{center}
\caption{\it Upper panels: Excluded region for
  $\text{BR}(\widetilde{t}\rightarrow t \widetilde\tau \tau)=1,0.8$ at
  the 90\% CL (left panel) and 95\% CL (right panel) in the plane
  $(m_{\widetilde{t}},m_{\widetilde{\nu}})$. For each value of the
  branching ratio the excluded region is the one enclosed by the
  corresponding curve. Above the thin dashed line the channel is
  kinematically forbidden. Middle panels: The same for
  $\text{BR}(\widetilde{t}\rightarrow t \widetilde\nu \nu) = 1, 0.8,
  0.6$. Lower panels: The same for $\text{BR}(\widetilde{t}\rightarrow
  b \widetilde\nu \tau) = 1,0.8,0.6,0.4$.}
\label{fig:bot}
\end{figure}

The results are reported in Fig.~\ref{fig:bot} where, for every given
channel, the bounds at the 90\%\, CL (left panels) and 95\%\,CL (right
panels) are presented in the plane $(m_{\widetilde
  t},m_{\widetilde\nu})$. Every panel contains the exclusion curves
corresponding to several values of the branching ratio into the
considered channel. For a given branching ratio, the allowed region
stands outside the respective curve (marked as in the legend) and
within the kinematically allowed area (below the thin dashed line).

For the decay $\widetilde t \to t \widetilde\tau \tau$ (upper
panels of Fig.~\ref{fig:bot}) the most sensitive analysis is the ATLAS
counting experiment. We combine it with the CMS signal region into a
single statistics. As Fig.~\ref{fig:bot} shows, the bound on this
channel is very weak. In particular, among the searches that we
identified as the most sensitive ones to this channel, there is no one
constraining this decay mode at 95\% CL for $m_{\widetilde t}\gtrsim
300\,$GeV and $m_{\widetilde\nu}\lesssim 100$\,GeV.

For the decay channel $\widetilde t \to t \widetilde\nu \nu$ (middle panels
of Fig.~\ref{fig:bot}) the most sensitive analysis is the CMS
analysis, though the ATLAS search for hadronically decayed stops is
also rather constraining. The bound provided in Fig.~\ref{fig:bot} is
based on the combination of both. As already pointed out, the
stringent cuts optimized for the searches for stops into on-shell LSP
neutralinos have rather low efficiency on the ``double invisible''
three-body decay signal involving an off-shell
mediator~\cite{Alves:2013wra}.

Finally, the bounds for the $\widetilde t \to b \widetilde\nu \tau$ decay channel are
presented in the lower panels of Fig.~\ref{fig:bot}. As summarized in
Tab.~\ref{tab:analyses}, it turns out that the most sensitive analysis
to this channel is the ATLAS counting one, although the other two
searches can also (slightly) probe this mode. In Fig.~\ref{fig:bot},
the exclusion curves for this channel are obtained by combining the
CMS signal regions with the ATLAS counting one into a single statistics (we do
not expect relevant improvements by also including the excluded ATLAS
analysis).

We expect the findings to be qualitatively independent of the
particular SUSY realization we consider.  The only model dependence is
the mass splitting between the stau and the sneutrino, which
determines the kinematic distribution of the stau decay products. In
specific SUSY models such a splitting is determined, and due to the
numerical approach of the present analysis, our results are obtained
for a concrete stau-sneutrino mass splitting, as detailed in
Sec.~\ref{sec:constraints}. Nevertheless, in practice, our results
should qualitatively apply to all SUSY realizations with prompt decays
of staus with mass $m_{\widetilde \tau} \lesssim m_{\widetilde
  \nu}+30\,$GeV and $BR(\widetilde \tau\to \widetilde \nu W^*)\simeq
100\%$~\footnote{To clarify this issue, we repeated the $\widetilde t
  \to t \widetilde\tau \tau$ simulations for a few parameter points
  featuring a tiny stau sneutrino mass splitting. For these few
  points, the constraints on $\widetilde t \to t \widetilde\tau \tau$
  presented in this paper turn out to be comparable, i.e.~ruling
  out a similar region of the parameter space in the plane
  $(m_{\widetilde t},m_{\widetilde\nu})$. Moreover the constraints on
  $\widetilde t \to t \widetilde\nu \nu$ and $\widetilde t \to b
  \widetilde\nu \tau$ are of course the same. This suggests that the
  presented bounds can be applied to other scenarios. Extensive
  parameter space simulations would be however required to prove this
  feature in full generality.}.

\subsection{Combined bounds}
\label{sec:comb}

In concrete models, it is feasible that the branching ratios of the
three aforementioned stop decay channels sum up to essentially 100\%, as
we will explicitly see in Sec.~\ref{sec:constraints}. In such a situation,
we can consider $\text{BR}(\widetilde{t}\rightarrow t
\widetilde{\nu}\nu)$ and $\text{BR}(\widetilde{t}\rightarrow b
\widetilde{\nu}\tau)$ as two independent variables, and fix
$\text{BR}(\widetilde{t}\rightarrow t \widetilde{\tau}\tau)$ as
\be
\label{eq:BRstautau}
\text{BR}(\widetilde{t}\rightarrow t
\widetilde{\tau}\tau)=1-\text{BR}(\widetilde{t}\rightarrow t
\widetilde{\nu}\nu)-\text{BR}(\widetilde{t}\rightarrow b
\widetilde{\nu}\tau)~.
\ee
It is then possible to use the aforementioned ATLAS and CMS searches
to constrain the two-dimensional plane
$\left[\text{BR}(\widetilde{t}\rightarrow t \widetilde{\nu}\nu),
\text{BR}(\widetilde{t}\rightarrow b \widetilde{\nu}\tau)\right]$ for some
set of values of $m_{\widetilde{t}}$ and $m_{\widetilde{\nu}}$.

The total number of signal events after cuts is given by
\be
  \label{eq:N}
  N = \sum_{i,j} N_{ij}(m_{\widetilde{t}})~
  \epsilon_{ij}(m_{\widetilde{t}},m_{\widetilde{\nu}})~,
\ee
with
\be
N_{ij}(m_{\widetilde{t}})=\mathcal{L}\,\sigma(p p\rightarrow
  \widetilde{t}\widetilde{t}^*)\times
  \text{BR}(\widetilde{t}\rightarrow i)\times
  \text{BR}(\widetilde{t}^*\rightarrow j)~,
\end{equation}
where $\mathcal{L} = 13$ fb$^{-1}$ stands for the integrated
luminosity, $\sigma$ is the stop pair production cross section, and
the indices $i$ and $j$ run over the three decay modes. The quantity
$\epsilon_{ij}$ is the efficiency that our recast analyses have on the
$\widetilde t \widetilde t^*\to ij$ events and is strongly dependent
on the mass spectrum. To determine $\epsilon_{ij}$ in some given mass
spectrum scenarios, we run simulations of $\widetilde t \widetilde
t^*\to ij$ following the procedure discussed above. As the searches do
not discriminate between $ij$ and its hermitian conjugate, it holds
$\epsilon_{ij}=\epsilon_{ji}$.

\begin{figure}[!h]
\begin{center}
 \includegraphics[width=0.49\columnwidth]{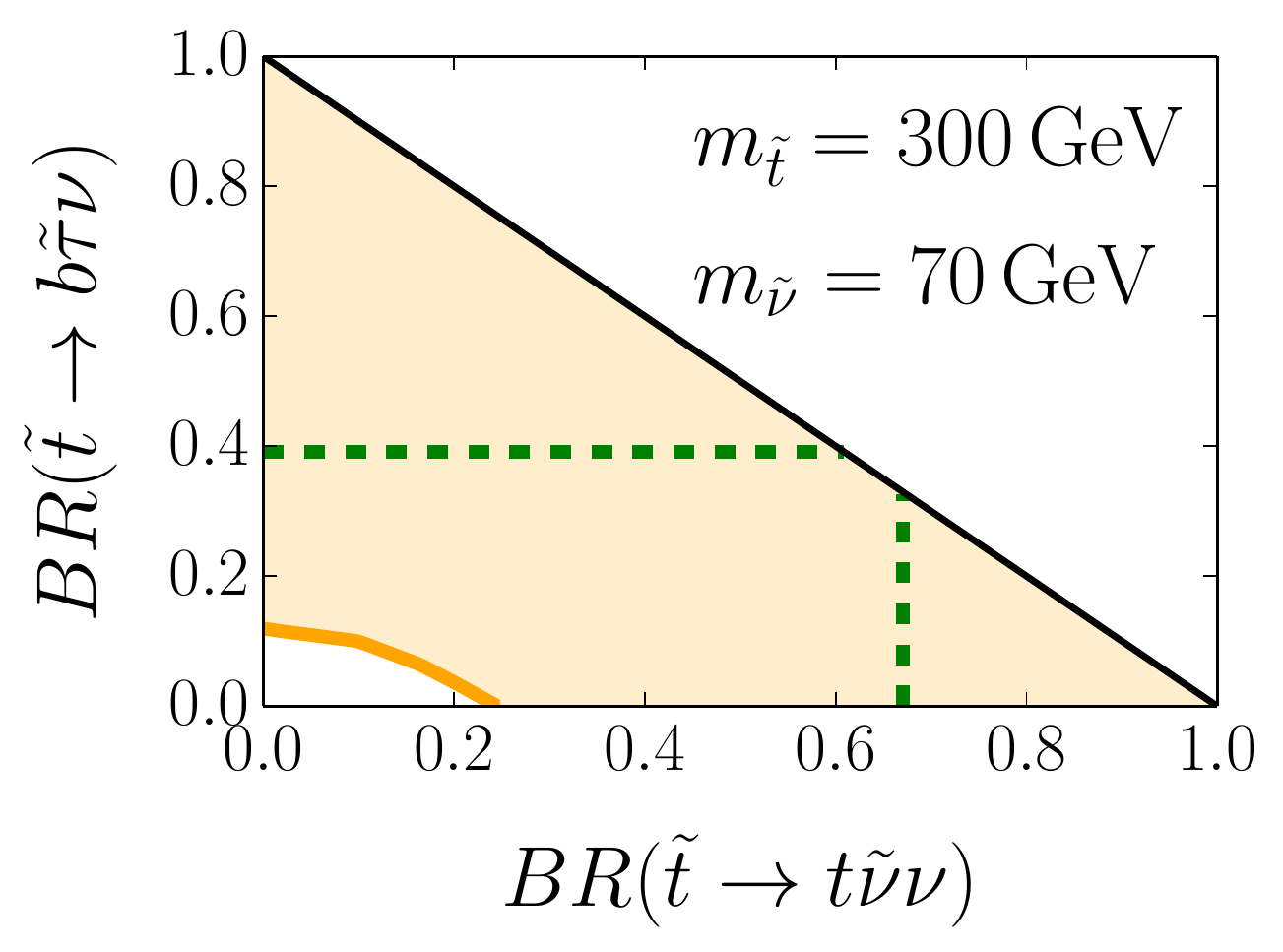}
 \includegraphics[width=0.49\columnwidth]{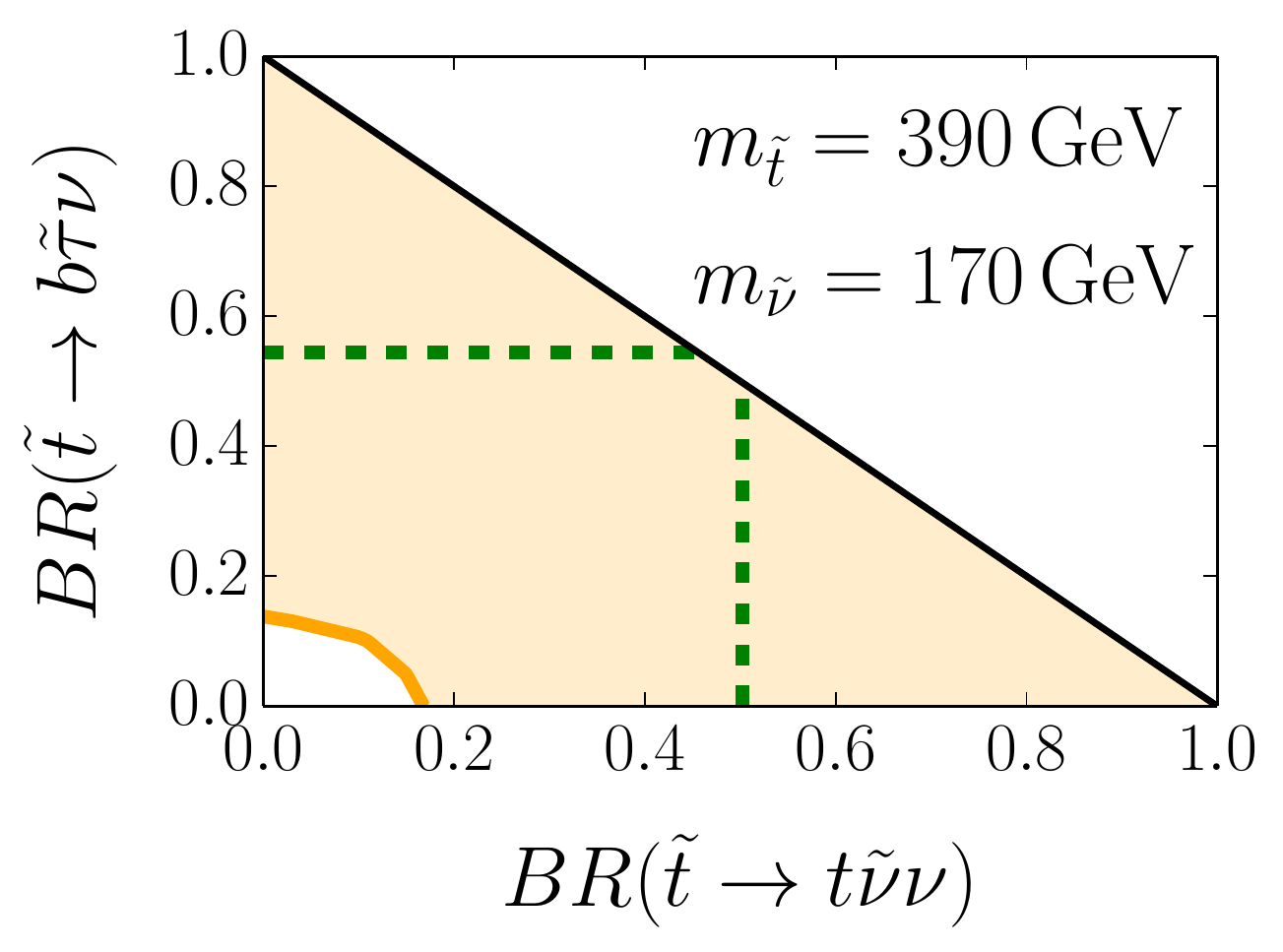}
 \includegraphics[width=0.49\columnwidth]{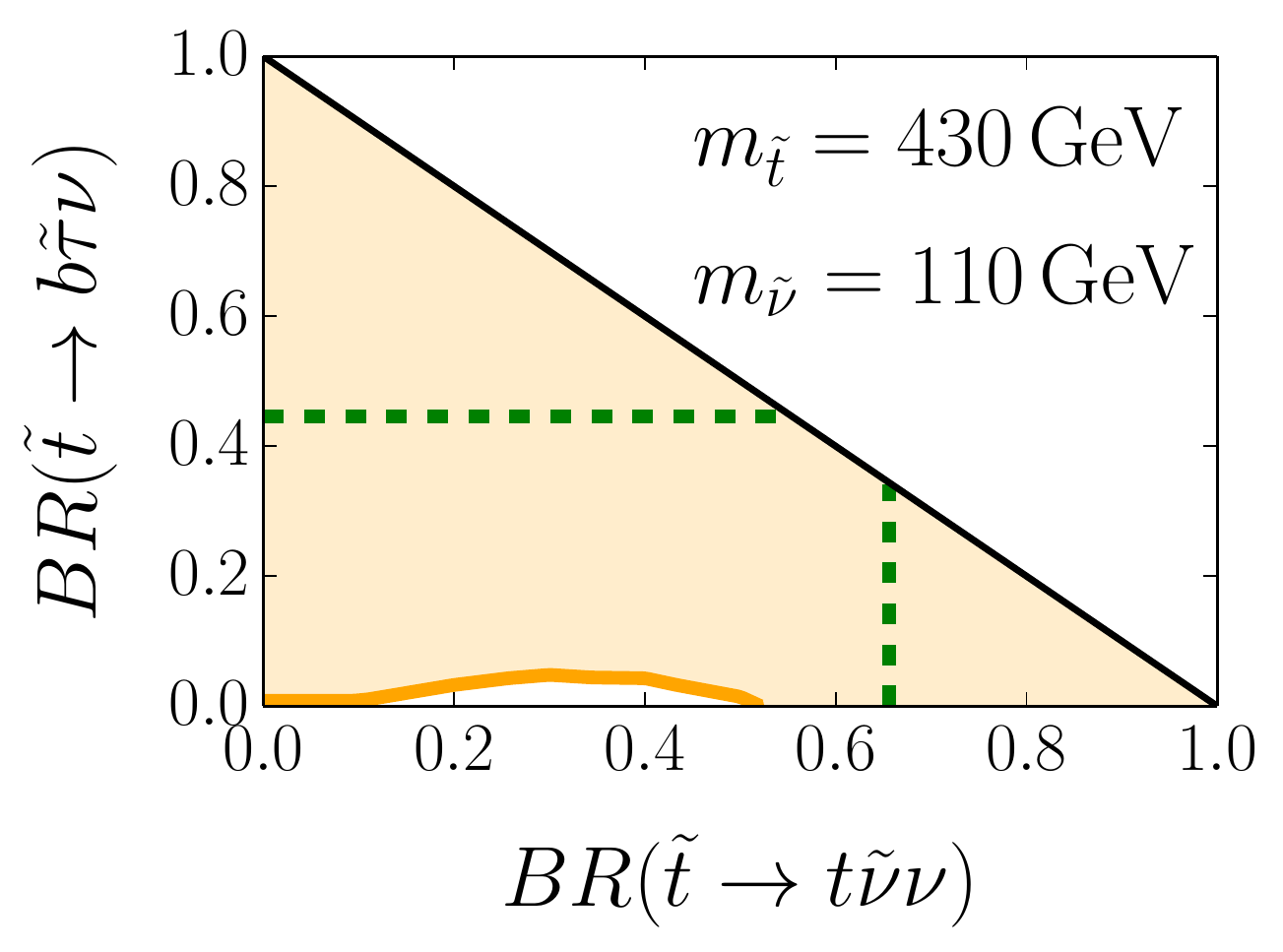}
 \includegraphics[width=0.49\columnwidth]{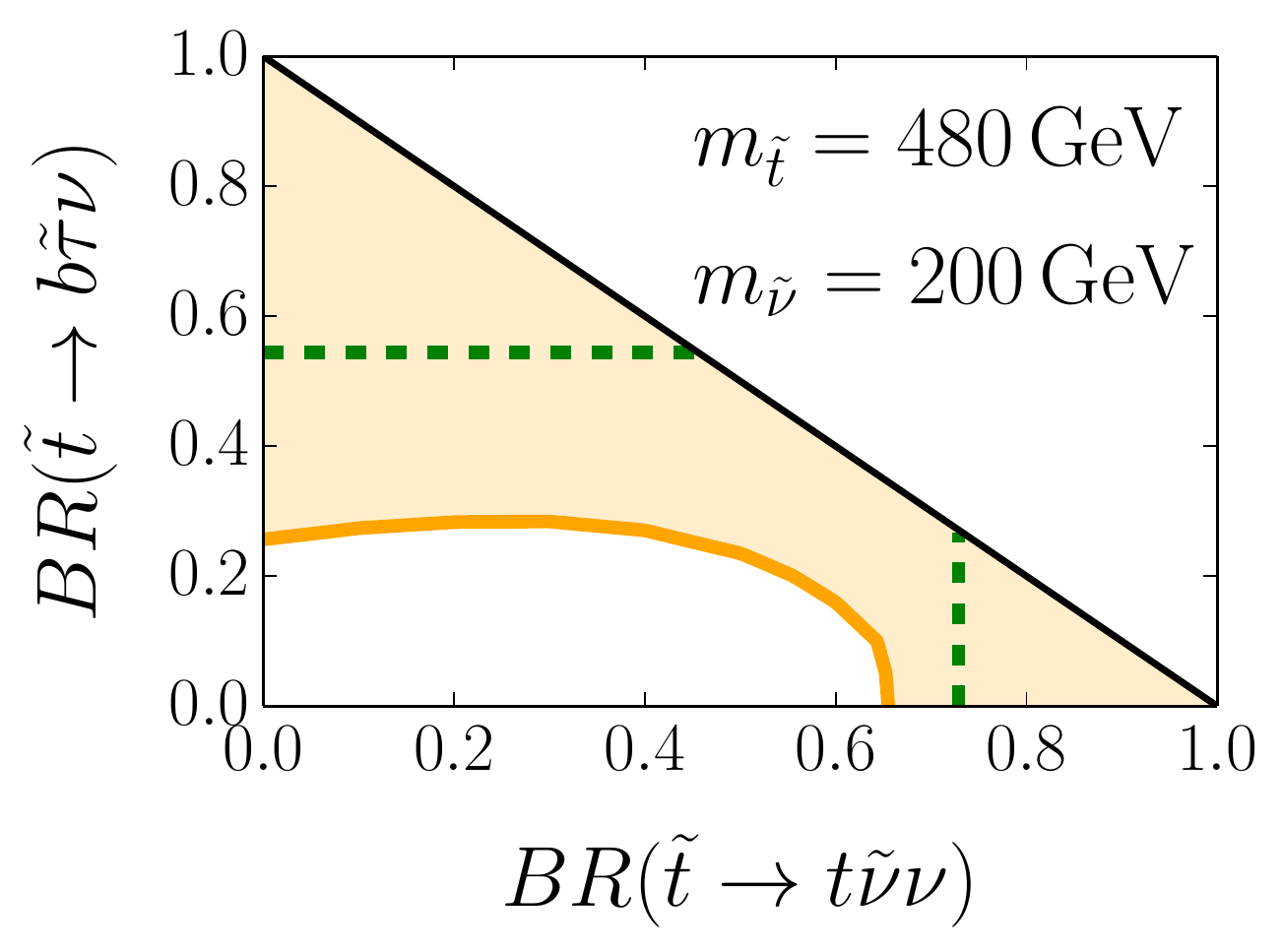}
\end{center}
\caption{\it Excluded regions at 95\% CL in the plane of BRs for
  different pairs of $(m_{\widetilde{t}}, m_{\widetilde{\nu}})$. The
  areas below (to the left of) the horizontal (vertical) green dashed
  lines would be allowed if only the $\widetilde t \to b\widetilde{\tau}\nu$ ($\widetilde t \to t\widetilde{\nu}\nu$) mode was considered. The areas enclosed by the orange solid lines are excluded when
  all channels are combined. The areas above the diagonal black solid straight lines are forbidden by the condition of Eq.~(\ref{eq:BRstautau}).}\label{fig:combined}
\end{figure}

The results are shown in Fig.~\ref{fig:combined}. The regions above
the horizontal dashed green lines would be the excluded ones had we
assumed the signal to consist of only $\widetilde t \widetilde t^*\to
b\widetilde{\nu}\tau b\widetilde{\nu}\tau$ events. Analogously, the
areas to the right of the vertical green dashed lines would be the
excluded ones under the assumption that only the events $\widetilde t
\widetilde t^*\to t\widetilde{\nu}\nu t\widetilde{\nu}\nu$ are
bounded. The regions enclosed by the orange solid lines are instead
excluded considering the whole signal, including also the stop decay
into $t\widetilde{\tau}\tau$ and the mixed channels. For such
comprehensive exclusion bounds, a common CL$_s$ is constructed out of
the bins in the ATLAS signal region SRB,
all bins in the CMS analysis and the single bin in the ATLAS counting
experiment.

In light of these results, several comments are in order: 

\begin{itemize}
\item \textit{i)} The comprehensive bounds, which exclude the region
  outside the orange curves, are much stronger than those obtained by
  the simple superposition of the constraints on the isolated signals,
  ruling out the region above and on the right of the horizontal and
  vertical dashed lines, respectively. This even reaches points close
  to the origin, where the main decay channel is
  $\widetilde{t}\rightarrow t\widetilde{\tau}\tau$. The main reason is
  the inclusion of the mixed channels.

\item
\textit{ii)} The fact that no single decay necessarily dominates,
makes sizeable regions of the parameter space to still be allowed by current
data. This is further reinforced by the smaller efficiencies that
current analyses have on these processes in comparison to the standard
channels. Thus, even small masses such as $m_{\widetilde{t}} \simeq
300$\,GeV and $m_{\widetilde{\nu}} \simeq 70$\,GeV, illustrated in the
top left panel, can be allowed.

\item \textit{iii)} As we can see from all panels in
  Fig.~\ref{fig:combined}, the allowed regions favor large values of
  $\text{BR}(\widetilde{t}\rightarrow t \widetilde{\tau}\tau)$.  This
  effect can be easily understood from the first row plots in
  Fig.~\ref{fig:bot}: \textit{there is little sensitivity of the present
  experimental searches to the channel} $\widetilde{t}\rightarrow t
  \widetilde{\tau}\tau$ when $m_{\widetilde t}$ and $m_{\widetilde
    \nu}$ are small.
\end{itemize}

\section{Constraints on particular SUSY models}
\label{sec:constraints}

The results of Sec.~\ref{sec:analyses} can be reinterpreted in
concrete SUSY scenarios that exhibit stops decaying as in
Fig.~\ref{fig:decays}, at least at detector scales~\footnote{As the Higgs plays no
role in this study, the origin of electroweak
breaking remains generically unspecified and not used to constrain the SUSY
parameters.}.
\begin{table}[htb]
 \begin{center}
\begin{adjustbox}{width=0.5\textwidth}
\footnotesize
\begin{tabular}{||l|c|c|c||}\hline
\textit{Scenario} & $M_1$ & $M_2$ & $\mu$\\  
\hline
 A&1.1 TeV &5 TeV &5 TeV  \\
 B & 1.1 TeV& 1.1 TeV& 1.1 TeV \\
 \hline
 \end{tabular}
 \end{adjustbox}
 \end{center}
 \caption{\it The value of the electroweakino mass parameters assumed
   in scenarios A and B.}\label{tab:scenarios}
\end{table}
The stop, stau, sneutrino and electroweakino mass spectrum and their
partial widths are determined by means of \texttt{SARAH v4} and
\texttt{SPheno v3}.  More specifically, we use the MSSM implementation
provided by these codes, and fix the parameters as follows.  We impose
$\tan\beta=10$, in agreement with the arguments of
Sec.~\ref{sec:model}. The slepton and squark soft-breaking trilinear
parameters are set to zero. The soft masses of the RH stop,
$M_{U_R}^2$, and LH stau doublet, $M_{L_L}^2$,
are much lighter than those of their partners with opposite
``chirality'', $M_{Q_L}^2$ and $M_{E_R}^2$. The electroweakino soft parameters are set, as shown for scenarios A and B in Tab.~\ref{tab:scenarios}, above the
lighter stop mass. The masses of the remaining SUSY particles are not
relevant for our analysis, they just need to be heavy enough to not
intervene in the stop phenomenology. Nevertheless, for practical
purposes, all SUSY parameter have to be chosen and then we set all
masses of the SUSY particles except electroweakinos, light stop and
light stau doublet at 3\,TeV.

\begin{figure}[htb]
\begin{center}
  \includegraphics[width=0.32\columnwidth]{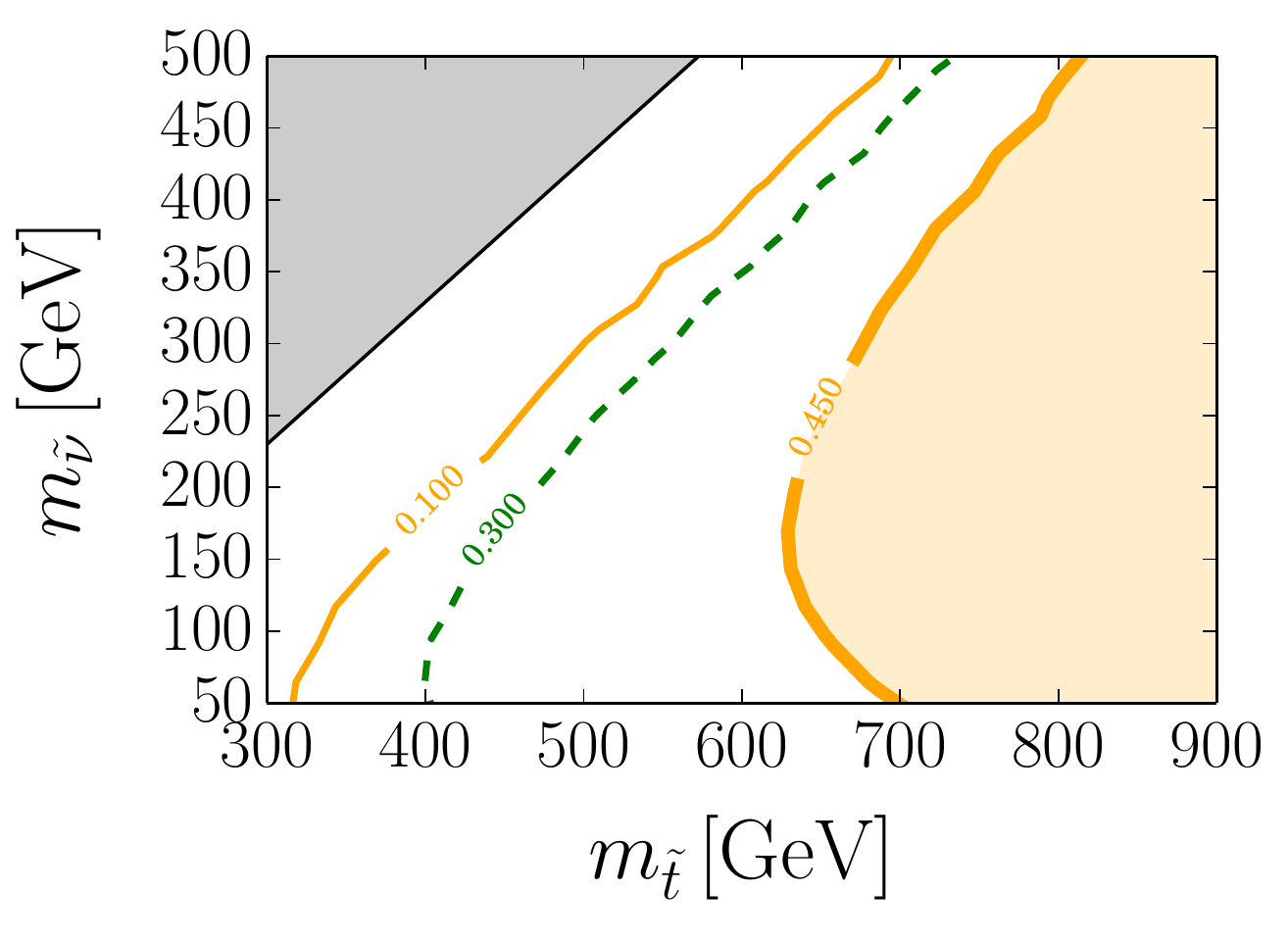}
  \includegraphics[width=0.32\columnwidth]{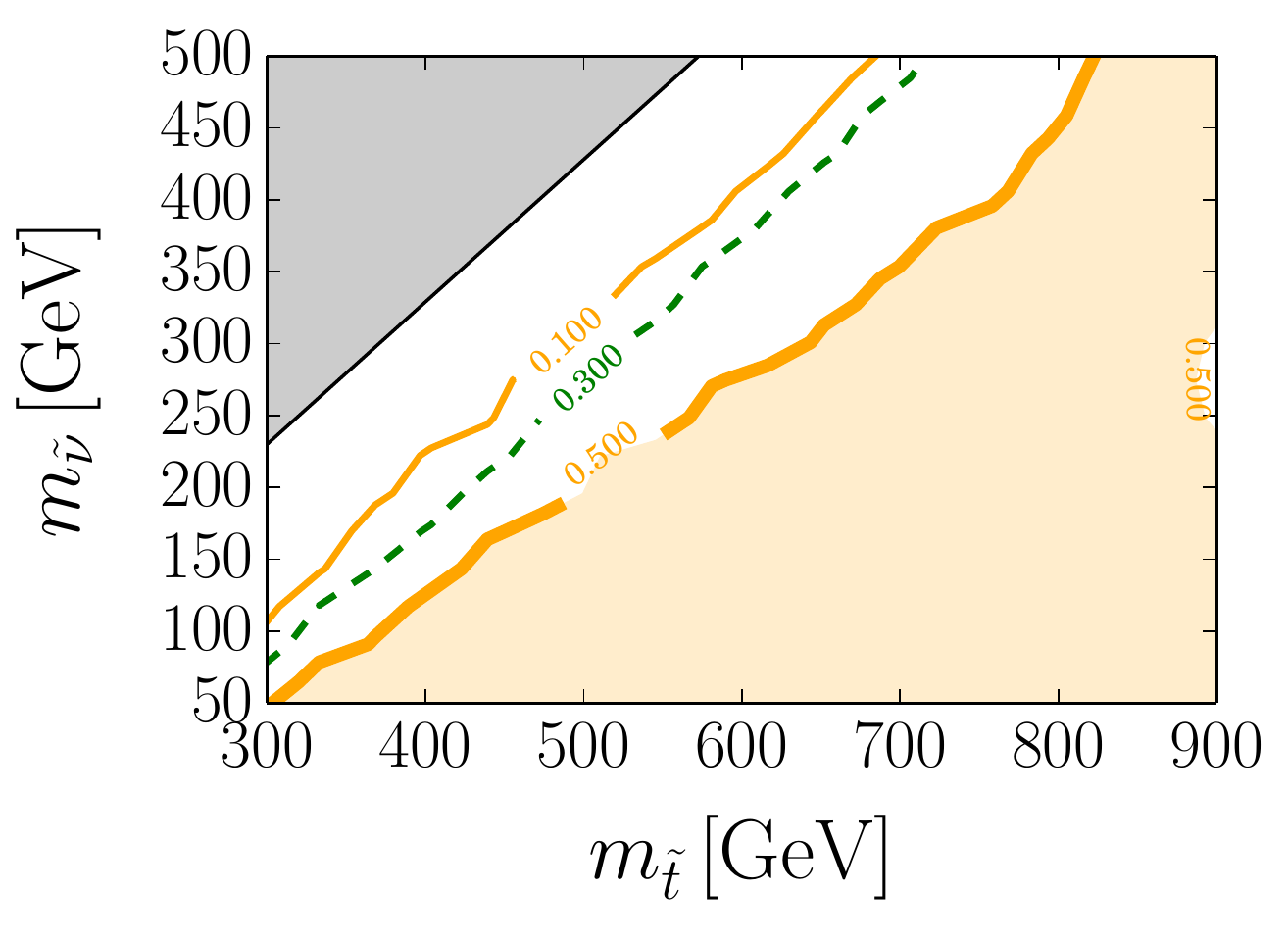}
  \includegraphics[width=0.32\columnwidth]{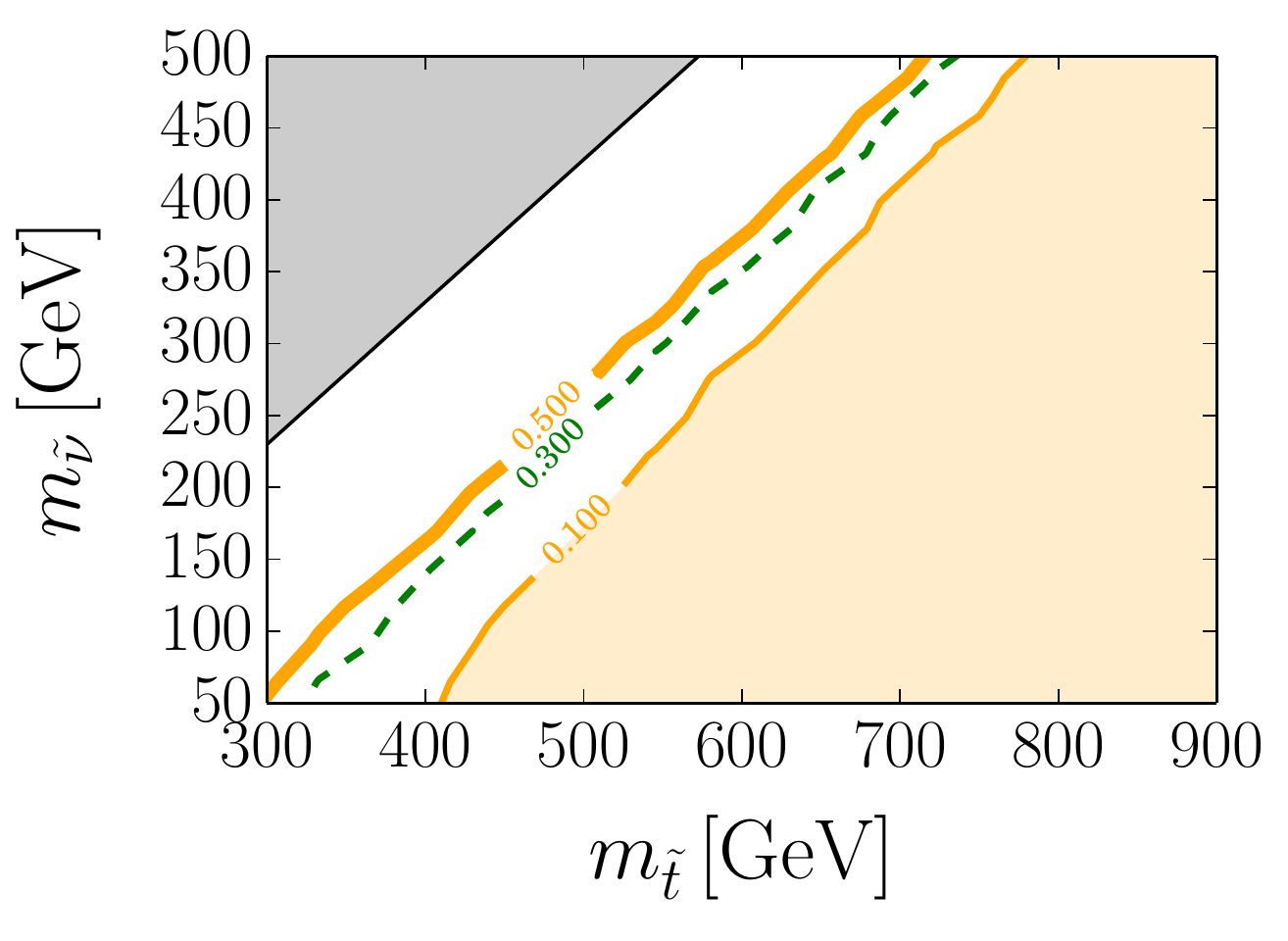}
  \includegraphics[width=0.32\columnwidth]{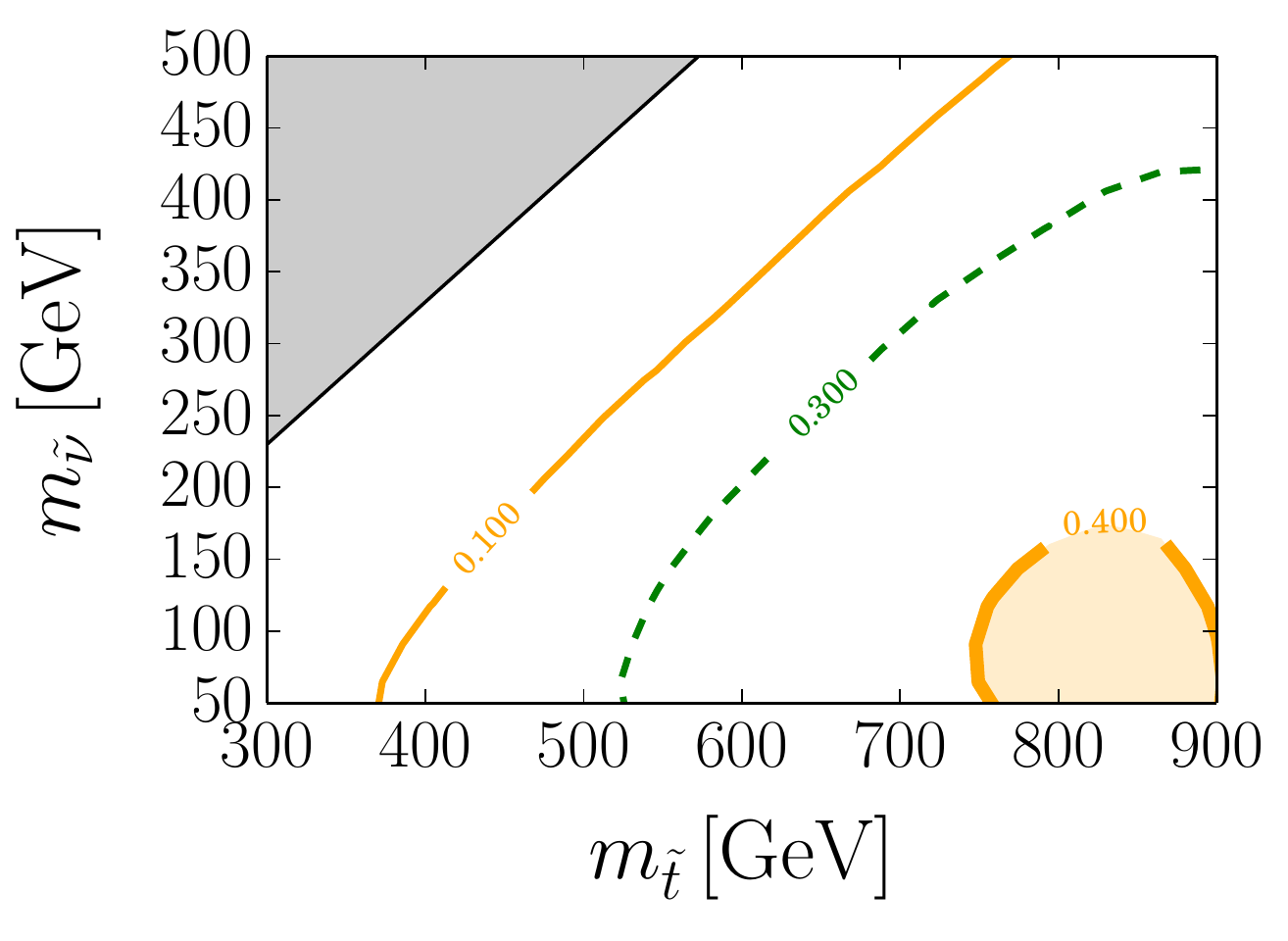}
  \includegraphics[width=0.32\columnwidth]{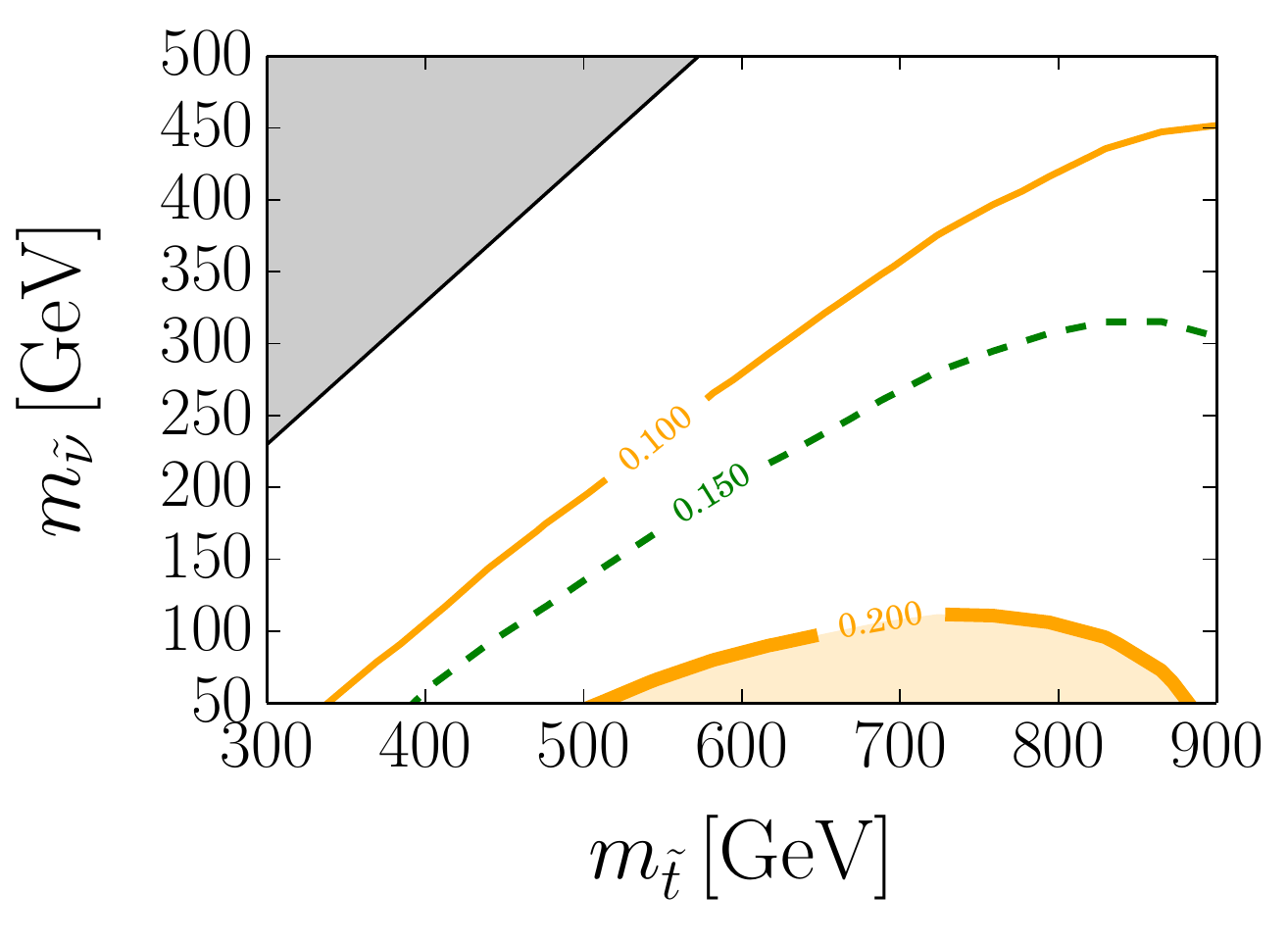}
  \includegraphics[width=0.32\columnwidth]{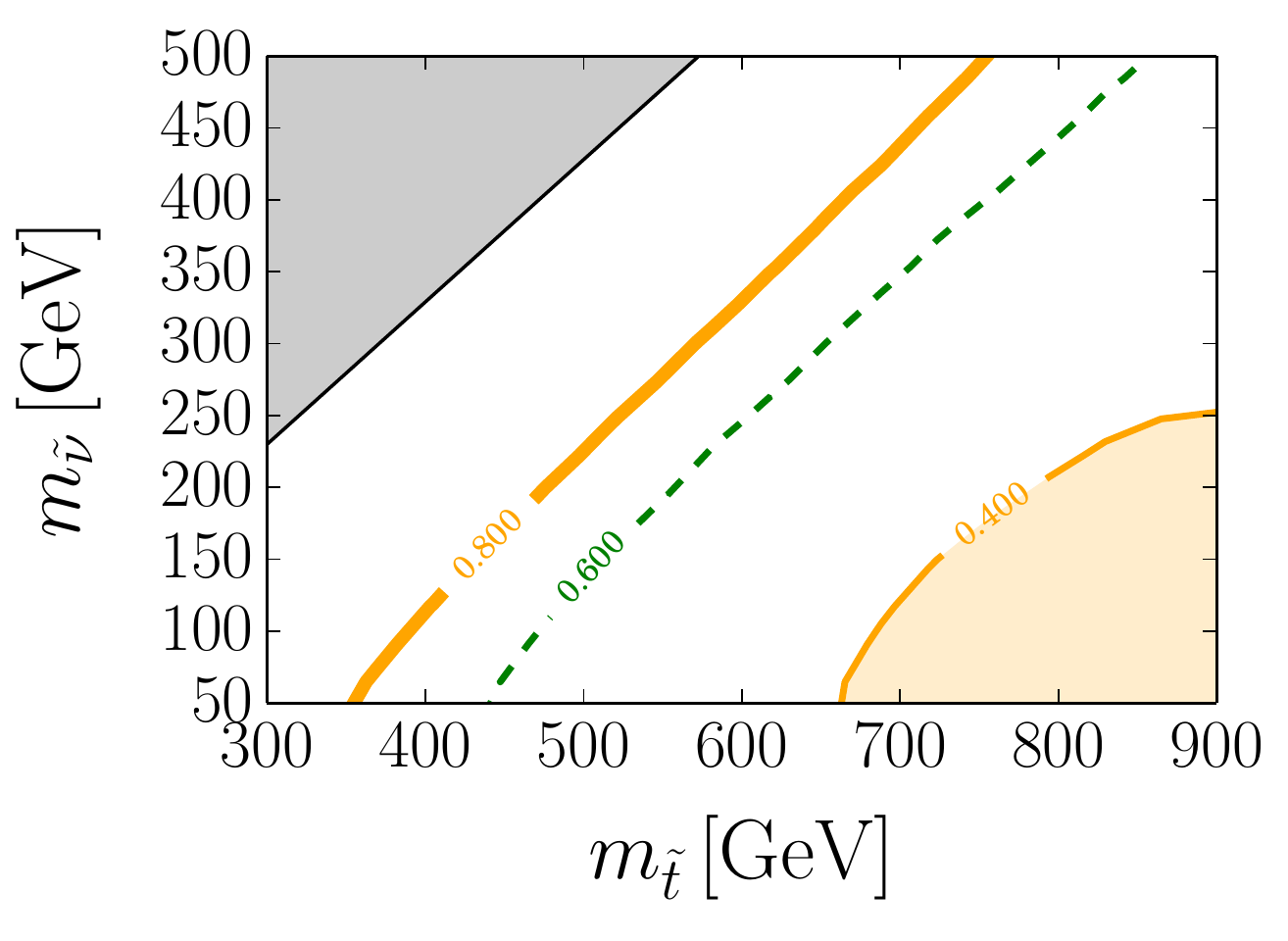}
\end{center}
\caption{\it Contour plots of the values of BR$(\widetilde{t}\to t
  \widetilde\tau \tau)$ (left panels), BR$(\widetilde{t}\to t
  \widetilde\nu \nu)$ (middle panels) and BR$(\widetilde{t}\to b
  \widetilde\nu \tau)$ (right panels) in Scenario A (upper panels) and Scenario
  B (lower panels).}
  \label{fig:BR}
\end{figure}

For the above parameter choice, we study two parameter regimes denoted
as scenarios A and B, characterized by the values of $M_1$,
$M_2$ and $\mu$ quoted in Tab.~\ref{tab:scenarios}. Within each
regime, we vary the masses $m_{\widetilde t}$ and $m_{\widetilde \nu}$,
by scanning over $M_{U_R}^2$ and $M_{ L_L}^2$,
and consequently $m_{\widetilde \tau}$ is determined as well. We
discard the parameter points with $m_{\widetilde t}< m_{\widetilde
  \nu}+70\,$GeV, which correspond to compressed scenarios that are
not investigated in this paper.  Contour plots of dominant stop
branching ratios are plotted in Figs.~\ref{fig:BR} as a function of
$m_{\widetilde t}$ and $m_{\widetilde \nu}$, for scenario A (upper row
panels) and scenario B (lower row panels). For each scenario, the
branching ratios of $\widetilde t\to t \widetilde\tau \tau$,
$\widetilde t\to t\widetilde\nu\nu$, and $\widetilde t\to
b\widetilde\nu \tau$ are plotted in the left, middle and right panels,
respectively.  As anticipated in Sec.~\ref{sec:model}, the main effect
of decreasing $M_2$ and $\mu$ is to enhance $BR(\widetilde t \to b
\widetilde\nu \tau)$, as we can see by comparing the two right panels
in Fig.~\ref{fig:BR}. Conversely, by increasing the value of $M_2$ and
$\mu$ we increase the branching ratio corresponding to the channel
$t\widetilde\tau\tau$, and we expect to make softer the bounds in the plane
$(m_{\widetilde t},m_{\widetilde\nu})$, in agreement with the
general behavior in the lower row panels in Fig.~\ref{fig:bot} and in
all plots in Fig.~\ref{fig:combined}. We stress that, within the
considered parameter range, the sum of these three branching ratios is
always above 95\% (depending on the range of $m_{\widetilde t}$ and
$m_{\widetilde\nu}$) which is consistent with our general model
assumptions.  We also checked numerically that the total width of the
stau is $\mathcal O(10^{-8}\,$GeV) for $m_{\widetilde \nu}\approx
500\,$GeV, and is much larger at smaller sneutrino
masses. Analogously, the mass gap between the stau and sneutrino masses
ranges between $5-40$ GeV, the latter value appearing for
$m_{\widetilde \nu}\approx 60\,$GeV.

\begin{figure}[h]
\begin{center}
 \includegraphics[width=0.49\columnwidth]{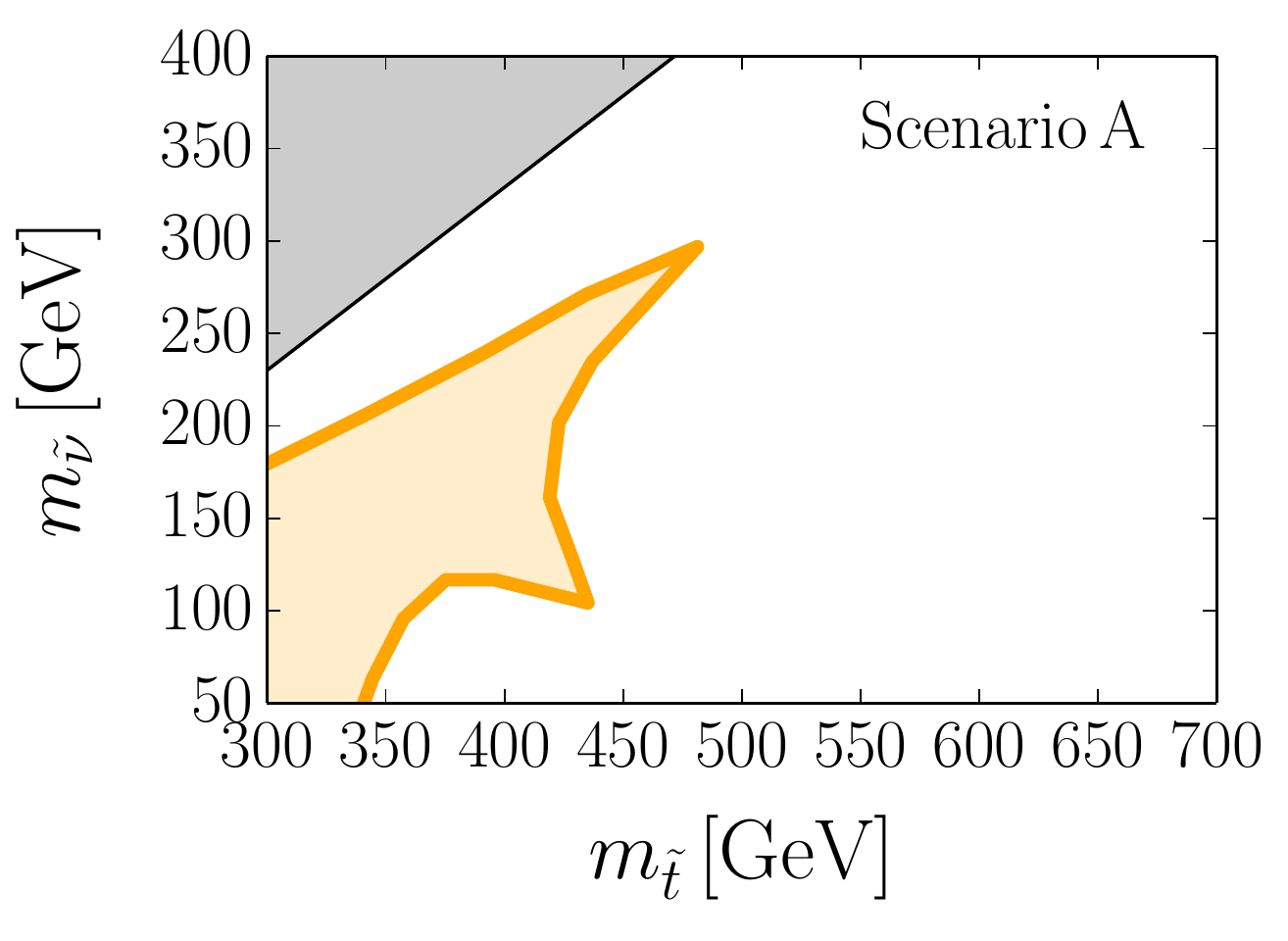}
 \includegraphics[width=0.49\columnwidth]{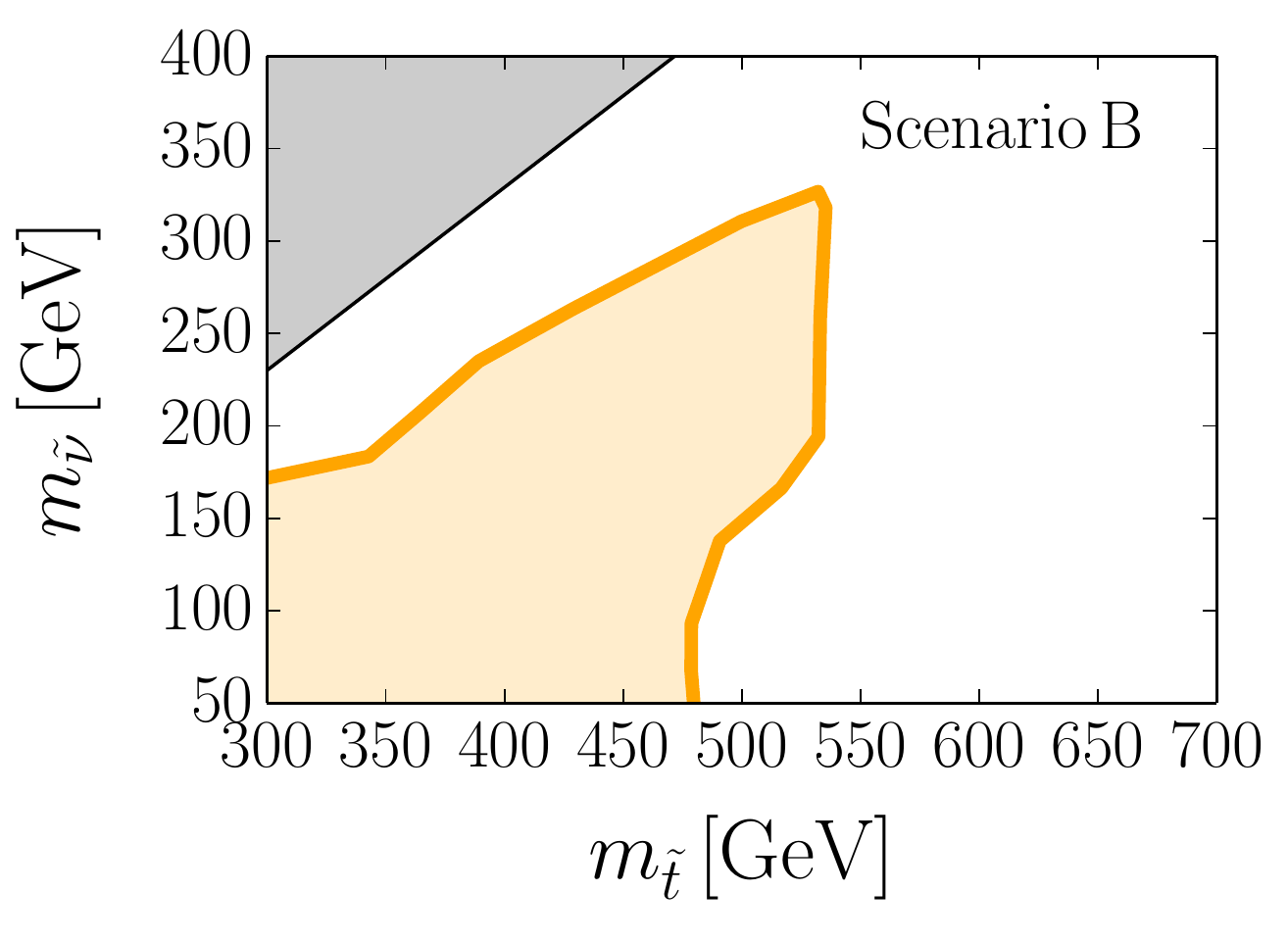}
 \end{center}
\caption{\it 95\% CL exclusion plots (inside the orange lines) for
  scenario A (left panel) and scenario B (right panel). The gray areas
  correspond to the region with $m_{\widetilde t}< m_{\widetilde
    \nu}+70\,$GeV that we do not investigate.}
  \label{fig:mssm}
\end{figure}
The results of Sec.~\ref{sec:analyses}, along with the numerical
evaluations of the different stop branching ratios, allow to recast
the present LHC constraints on scenarios A and B.  At each parameter
point we rescale the amount of signal events, depending on the values
of the branching ratios extracted from the MSSM parameter card
corresponding to that point~\footnote{In order to check the
  consistency of our procedure, we also perform the collider
  simulations described in Sec.~\ref{sec:analyses} for numerous
    parameter configurations of each scenario. We find perfect
  consistency, meaning that the contribution from any channel to the
  search of any other is negligible.}. The final excluded regions at
95\% CL in the plane $(m_{\widetilde t},m_{\widetilde\nu})$ are shown
in Fig.~\ref{fig:mssm}. Both in scenario A (left panel) and B (right
panel) the exclusion bounds (orange areas) are relaxed with respect to
their analogous in SUSY scenarios with the neutralino as the LSP.  As
anticipated, bounds are weaker in scenario A than in scenario B, due
the larger values of BR($\widetilde t\to
t\widetilde\tau\tau$). Remarkably, in the presence of light
sneutrinos, a RH stop at around 350\,GeV is not ruled out by current
LHC data, or at least by the ATLAS and CMS analyses performed till
now.

\section{Conclusions}
\label{sec:conclusions}

The bottom line in this paper is that, in the minimal supersymmetric
standard model (MSSM) scenario with heavy electroweakinos, light staus
and light tau sneutrinos, a mostly right-handed stop with a mass of
around 350\,GeV is compatible with the present LHC data. This is
mostly due to the coexistence of several branching ratios into
channels which the LHC searches have weak sensitivity to.  Although we
have not been concerned about detailed naturalness issues, light stops
certainly help in this sense. Heavy electroweakinos are instead
considered unnatural, but this is not necessarily true for low scale
supersymmetric (SUSY) breaking. In particular, heavy electroweakinos
are feasible without inducing a hierarchy problem in some
supersymmetry breaking embeddings based on Scherk-Schwarz (SS) and low
scale gauge mediated supersymmetry breaking (GMSB) mechanisms.

In the investigated scenario, the light spectrum only includes the
Standard Model particles, the mostly right-handed stop, the tau sneutrino and the
mostly left-handed stau. Among these SUSY particles, the light stop is
heavier than the left-handed stau, which is in turn heavier than the
tau sneutrino. The charginos and neutralinos might be at the TeV scale
or below, but in any case heavier than the light stop. The number of
dominant stop decay channels is only a few. These decays occur via
off-shell electroweakinos, and ATLAS and CMS fully hadronic searches
for stops into hadronic or tau lepton
states~\cite{atl:2016sab,cms:2016srd,atl:2016src}, although designed
for a different scope, are the searches that are expected to be most
sensitive to them. Remarkably, their constraints do not rule out stops
with masses as small as 350\,GeV, when the stau mass is around
100\,GeV, the sneutrino mass is approximately 60\,GeV, and the
electroweakinos are at the TeV scale. Neither further bounds do apply:
such staus are heavy enough to be compatible with the LEP
bounds~\cite{Barate:1997dr,Abreu:2000tn,Abbiendi:2003yd,Achard:2001qw},
and decay fast, in agreement with the LHC bounds on disappearing
tracks~\cite{Kopeliansky:2015gbi,Khachatryan:2016sfv}.

The only constraint comes from cosmological scale observables. In the
present study the tau sneutrino is the lightest SUSY
particle, stable (at least) at collider scales. If it also is stable
at cosmological scales, its thermal relic density is below the dark
matter (DM) abundance~\cite{Falk:1994es,Arina:2007tm} and, moreover, it is
also ruled out by direct detection
experiments~\cite{Tan:2016zwf,Akerib:2016vxi}. So the scenario has to
be completed somehow, to provide a reliable explanation of the
surveyed DM relic density and/or avoid the strong bounds from
direct detection experiments.

There are a limited number of possible mechanisms to circumvent the
previous problems without altering the stop phenomenology we have
investigated. The simplest possibility is to assume that the
sneutrino, even though stable at collider scales, is \textit{unstable
  at cosmological scales}. In theories with $R$-parity conservation
this can be realized only if there is a lighter SUSY particle
(possibly a DM candidate) which the sneutrino decays to, but such that
the sneutrino only decays outside the detector and in cosmological
times. In theories with GMSB this role can be played by a light
gravitino $\widetilde G$. It is a candidate to warm DM and its
cosmological abundance is given by $\Omega_{3/2}h^2\simeq 0.1
(m_{3/2}/0.2 \textrm{ keV})$, which suggests a rather low scale of
supersymmetry breaking $F\simeq m_{3/2}M_P$. In this case the
sneutrino decays as $\widetilde\nu\to \nu\widetilde G$ and, as far as
collider phenomenology is concerned, it looks stable. In theories with
a heavy gravitino, as e.g.~in theories with SS breaking, one could
always introduce a right-handed sneutrino $\nu_R$, lighter than the
left-handed sneutrino~\footnote{This can be achieved for instance by
  localizing the right-handed neutrino multiplet in the brane and thus
  receiving its mass from higher order radiative corrections.}. On the
other hand, the right-handed sneutrino can in principle play the role
of DM~\cite{Cerdeno:2008ep,Arina:2015uea}. If its fermionic partner is
light, also the decay $\widetilde t\to b \widetilde \tau \nu_R$
appears although this process is suppressed by the small neutrino
Yukawa coupling. Thus, in practice, the stop collider
phenomenology would not be different from that considered in the present
paper. Another possibility is if the cosmological model becomes
non-standard, as would happen by assuming modifications of general
relativity or with non-standard components of DM, as for instance
black holes~\footnote{For discussions in this direction see
  e.g.~Refs.~\cite{Capela:2012uk,
    Clesse:2016vqa,Bird:2016dcv,McGaugh:2016leg}.}. In this case, in
order to overcome the direct detection bounds, the initial density of
sneutrinos in thermal equilibrium should be diluted by some mechanism,
as e.g.~an entropy production (or simply a non-standard expansion of
the universe), before the big bang
nucleosynthesis~\cite{Gelmini:2010zh, Nardini:2011hu}. Finally the simplest
solution to avoid the direct detection bounds is if there is a small
amount of $R$-parity breaking and the sneutrino becomes unstable at
cosmological scales. For instance one can introduce an $R$-parity
violating superpotential as $W=\lambda_{ijk}L_i L_j
E_k$~\cite{Dreiner:1997uz}, with a small Yukawa coupling
$\lambda_{ijk}$ such that the sneutrino decays as $\widetilde\nu\to
e_j \bar e_k$. Depending on the value of the coupling $\lambda$ the
sneutrino can decay at cosmological times. Needless to say, in this
case one would need some additional candidate to DM.

Remarkably, the present bounds on the stop mass in the considered
scenario are so weak that even the complete third-generation squarks
might be accommodated in the sub-TeV spectrum. Indeed, the kinematic
effects and the coexistence of multi decay channels responsible for
the poorly efficient current LHC searches, should also (partially)
apply to the left-handed third-family squarks. The presence of these
additional squarks in the light spectrum would effectively increase
the number of events ascribable to the channels we have
analyzed. Nonetheless, since the obtained constraints are very weak,
there should be room for a sizeable number of further events before
reaching TeV-scale bounds. In such a case, in the heavy electroweakino
scenario considered in this paper, present data could still allow for
a full squark third-family generation much lighter than what is
naively inferred from current constraints based on simplified
models. Quantifying precisely this, as well as studying the
right-handed neutrino extension, is left for future investigations.

Details aside, our main conclusion highlights the existence of unusual
scenarios where very light stops are compatible with the present LHC
searches without relying on artificial (e.g.~compressed) parameter
regions. It is not clear whether this simply occurs because of lack of
dedicated data analyses. \textit{In summary, the possibility that the
  bias for the neutralino as lightest SUSY particle have misguided the
  experimental community towards partial searches, and that clear SUSY
  signatures are already lying in the collected data, is certainly
  intriguing.}\\ 
  
\section*{ Acknowledgments}
  
\noindent
The work of MC is partially supported by the Spanish MINECO under grant FPA2014-54459-P
and by the Severo Ochoa Excellence Program under grant SEV-2014-0398.
The work of AD is partially supported by the National
Science Foundation under grant PHY-1520966.
The work of GN is supported by the Swiss National
Science Foundation under grant 200020-168988. The work of MQ is also partly supported
by the Spanish MINECO under grant CICYT-FEDER-FPA2014-55613-P, by the
Severo Ochoa Excellence Program under grant
SO-2012-0234, by Secretaria d'Universitats i Recerca del Departament
d'Economia i Coneixement de la Generalitat de Catalunya under grant
2014 SGR 1450, and by the CERCA Program/Generalitat de Catalunya.

\appendix

\section{Analysis validation}
\label{appendixA}
\begin{figure}[b]
\begin{center}
 \includegraphics[width=0.33\columnwidth]{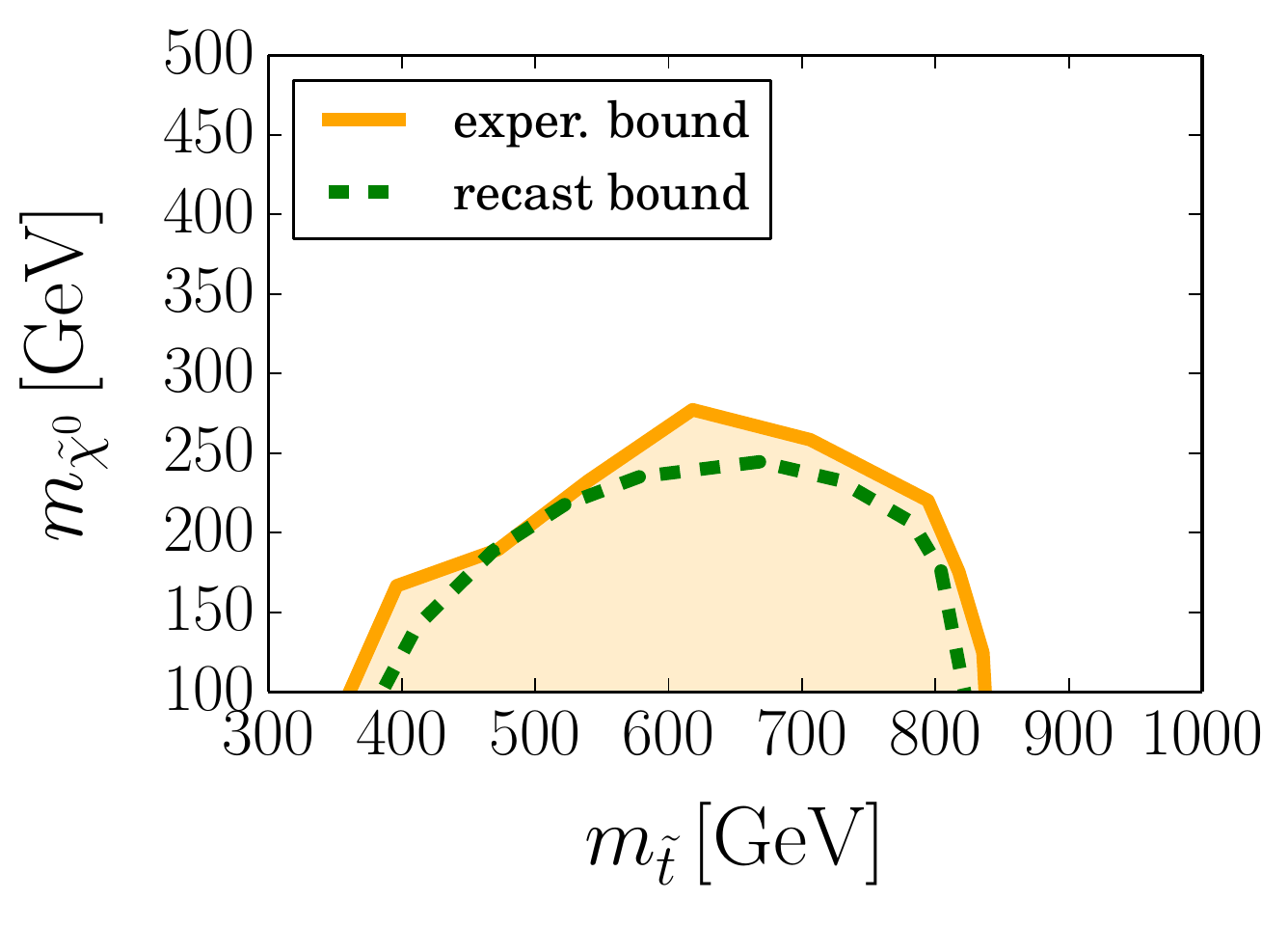}
    \includegraphics[width=0.33	\columnwidth]{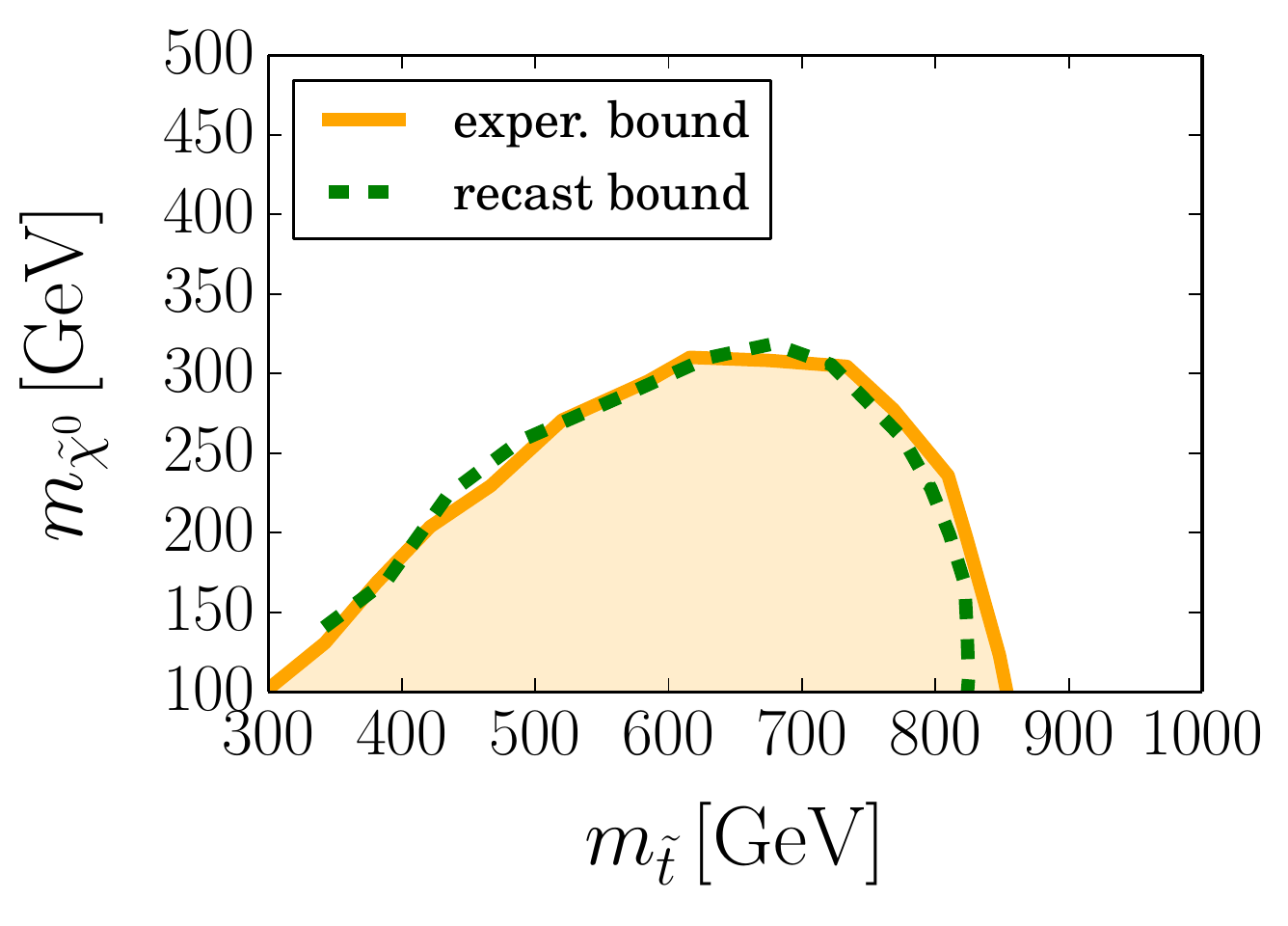}
    \includegraphics[width=0.33\columnwidth]{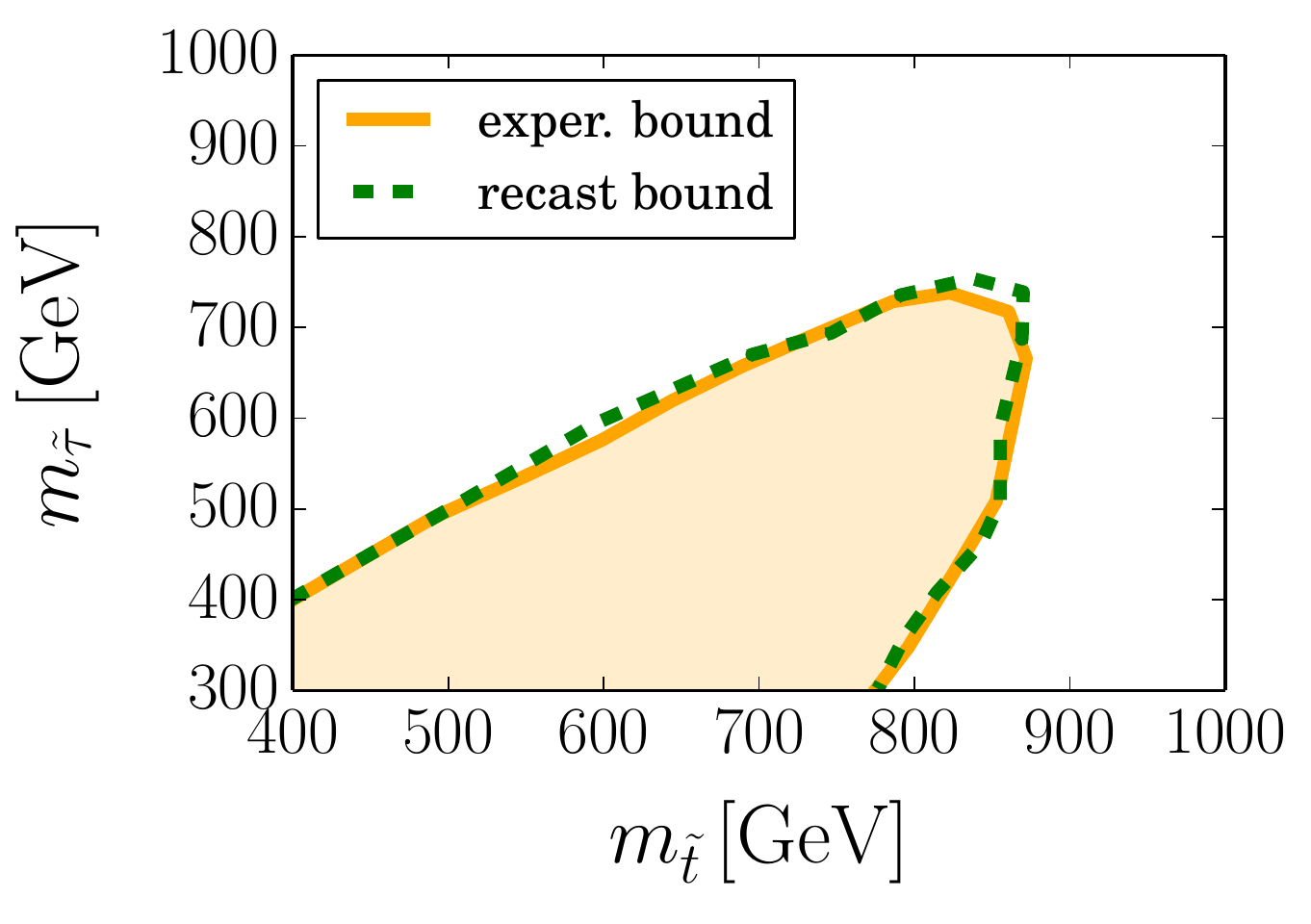}
\end{center}
\caption{\label{fig:validation} \it Comparison of the bounds reported
  by the experimental papers (solid orange lines) of
  Ref.~\cite{atl:2016sab} (left panel), Ref.~\cite{cms:2016srd}
  (middle panel) and Ref.~\cite{atl:2016src} (right panel) with those
  obtained after recasting the analyses (dashed green lines).}
\end{figure}
In order to validate our implementations of the experimental analyses
of Refs.~\cite{atl:2016sab, cms:2016srd, atl:2016src}, we apply them
to Monte Carlo events generated using the same benchmark models of
those searches.  Specifically, these are pair-produced stops decaying
as $\widetilde{t}\rightarrow t \widetilde{\chi}^0$~\cite{atl:2016sab,
  cms:2016srd} and $\widetilde{t}\rightarrow
b\overline{\nu}\widetilde{\tau}~(\widetilde{\tau}\rightarrow
\tau\widetilde{G})$~\cite{atl:2016src}. The signal samples are
obtained by generating pairs of stop events in the MSSM with
\texttt{MadGraph v5} at leading order. Such events are subsequently
decayed by \texttt{Pythia v6}. In the parameter cards produced with
\texttt{SARAH v4} and \texttt{SPheno v3}, the branching ratio
$\text{BR}(\widetilde{t}\rightarrow t \widetilde{\chi}^0)$ is fixed
manually to 100\% in the first two analyses.  In the same vein, for
the analysis of Ref.~\cite{atl:2016src} we fix both
$\text{BR}(\widetilde{t}\rightarrow b\overline{\nu}\widetilde{\tau}) =
1$ and $\text{BR}(\widetilde{\tau}\rightarrow \tau\nu) =
1$. Notice that, in this
last case, the neutrino plays the role of the (massless) gravitino,
thus mimicking the channel studied in the experimental work. As stated
in the main text, bounds are obtained by combining the different bins
of a particular search into a single statistics (note that the
analysis of Ref.~\cite{atl:2016src} is simply a counting
experiment). The only caveat concerns the analysis of
Ref.~\cite{atl:2016sab}. The two signal regions considered in that
search are not statistically independent.  Therefore, the most
constraining of the two statistics, each constructed out of the three
bins of a particular signal region, is taken.  Altogether, the
comparison between the bounds reported in Refs.~\cite{atl:2016sab,
  cms:2016srd, atl:2016src} and ours are displayed in
Fig.~\ref{fig:validation}. We have checked that QCD next-to-leading
order effects (taken as an overall K-factor) shift the dashed green
lines by only a small amount.
%



\begin{thebibliography}{99}



  


\bibitem{Barate:1997dr}
  R.~Barate {\it et al.} [ALEPH Collaboration],
  ``Search for pair production of longlived heavy charged particles in e+ e- annihilation,''
  Phys.\ Lett.\ B {\bf 405} (1997) 379
  [hep-ex/9706013].

\bibitem{Abreu:2000tn}
  P.~Abreu {\it et al.} [DELPHI Collaboration],
  ``Search for heavy stable and longlived particles in e+ e- collisions at s**(1/2) = 189-GeV,''
  Phys.\ Lett.\ B {\bf 478} (2000) 65
  [hep-ex/0103038].
\bibitem{Abbiendi:2003yd}
  G.~Abbiendi {\it et al.} [OPAL Collaboration],
  ``Search for stable and longlived massive charged particles in e+ e- collisions at s**(1/2) = 130-GeV to 209-GeV,''
  Phys.\ Lett.\ B {\bf 572} (2003) 8
  [hep-ex/0305031].
\bibitem{Achard:2001qw}
  P.~Achard {\it et al.} [L3 Collaboration],
  ``Search for heavy neutral and charged leptons in $e^{+} e^{-}$ annihilation at LEP,''
  Phys.\ Lett.\ B {\bf 517} (2001) 75
  [hep-ex/0107015].

\bibitem{Kopeliansky:2015gbi}
  R.~Kopeliansky,
  ``Search for charged long-lived particles with the ATLAS detector in pp collisions at $\sqrt{s}$ = 8 TeV,''
  CERN-THESIS-2015-119.
  
\bibitem{Khachatryan:2016sfv}
  V.~Khachatryan {\it et al.} [CMS Collaboration],
  ``Search for long-lived charged particles in proton-proton collisions at sqrt(s) = 13 TeV,''
  Phys.\ Rev.\ D {\bf 94} (2016) 112004
  [arXiv:1609.08382 [hep-ex]].



\bibitem{Bianchi:2242387} 
R.~ M.~Bianchi [ATLAS Collaboration], ``SUSY
  searches with the ATLAS detector'', ATL-PHYS-PROC-2017-005.

\bibitem{Kazana:2016gni}
  M.~Kazana [CMS Collaboration],
  ``Searches for Supersymmetry with the CMS Detector at the LHC,''
  Acta Phys.\ Polon.\ B {\bf 47} (2016) 1489.

\bibitem{Cerdeno:2008ep}
  D.~G.~Cerdeno, C.~Munoz and O.~Seto,
  ``Right-handed sneutrino as thermal dark matter,''
  Phys.\ Rev.\ D {\bf 79} (2009) 023510
  [arXiv:0807.3029 [hep-ph]].

\bibitem{Arina:2013zca}
  C.~Arina and M.~E.~Cabrera,
  ``Multi-lepton signatures at LHC from sneutrino dark matter,''
  JHEP {\bf 1404} (2014) 100
  [arXiv:1311.6549 [hep-ph]].

\bibitem{Arina:2015uea} C.~Arina, M.~E.~C.~Catalan, S.~Kraml,
  S.~Kulkarni and U.~Laa, ``Constraints on sneutrino dark matter from
  LHC Run 1,'' JHEP {\bf 1505} (2015) 142
  [arXiv:1503.02960 [hep-ph]].


\bibitem{Giudice:1998bp}
  G.~F.~Giudice and R.~Rattazzi,
  ``Theories with gauge mediated supersymmetry breaking,''
  Phys.\ Rept.\  {\bf 322} (1999) 419
  [hep-ph/9801271].
  
 
\bibitem{Delgado:2012rk}
  A.~Delgado, G.~Nardini and M.~Quiros,
  ``The Light Stop Scenario from Gauge Mediation,''
  JHEP {\bf 1204} (2012) 137
  [arXiv:1201.5164 [hep-ph]].

\bibitem{delAguila:1986klm}
  F.~del Aguila, M.~Quiros and F.~Zwirner,
  ``Detecting E(6) Neutral Gauge Bosons Through Lepton Pairs at Hadron Colliders,''
  Nucl.\ Phys.\ B {\bf 287} (1987) 419.

\bibitem{Langacker:2008yv}
  P.~Langacker,
  ``The Physics of Heavy $Z^\prime$ Gauge Bosons,''
  Rev.\ Mod.\ Phys.\  {\bf 81} (2009) 1199
  [arXiv:0801.1345 [hep-ph]].
 

  
  \bibitem{Scherk:1979zr}
J.~Scherk and J.~H. Schwarz, 
``How to Get Masses from Extra Dimensions,"
  Nucl.\ Phys. {\bf B153} (1979) 61--88.
  
\bibitem{Antoniadis:1990ew}
  I.~Antoniadis,
 `` A Possible new dimension at a few TeV,"
  Phys.\ Lett.\ B  {\bf 246} (1990) 377.

\bibitem{Pomarol:1998sd}
  A.~Pomarol and M.~Quiros,
  ``The Standard model from extra dimensions,''
  Phys.\ Lett.\ B {\bf 438} (1998) 255
  [hep-ph/9806263].

\bibitem{Antoniadis:1998sd}
  I.~Antoniadis, S.~Dimopoulos, A.~Pomarol and M.~Quiros,
  ``Soft masses in theories with supersymmetry breaking by TeV compactification,''
  Nucl.\ Phys.\ B {\bf 544} (1999) 503
  [hep-ph/9810410].

\bibitem{Delgado:1998qr}
  A.~Delgado, A.~Pomarol and M.~Quiros,
  ``Supersymmetry and electroweak breaking from extra dimensions at the TeV scale,''
  Phys.\ Rev.\ D {\bf 60} (1999) 095008
  [hep-ph/9812489].

\bibitem{Quiros:2003gg}
  M.~Quiros,
  ``New ideas in symmetry breaking,''
  hep-ph/0302189.

\bibitem{Dimopoulos:2014aua}
  S.~Dimopoulos, K.~Howe and J.~March-Russell,
  ``Maximally Natural Supersymmetry,''
  Phys.\ Rev.\ Lett.\  {\bf 113} (2014) 111802
  [arXiv:1404.7554 [hep-ph]].

\bibitem{Garcia:2015sfa}
  I.~Garcia Garcia, K.~Howe and J.~March-Russell,
  ``Natural Scherk-Schwarz Theories of the Weak Scale,''
  JHEP {\bf 1512} (2015) 005
  [arXiv:1510.07045 [hep-ph]].

\bibitem{Delgado:2016vib}
  A.~Delgado, M.~Garcia-Pepin, G.~Nardini and M.~Quiros,
  ``Natural Supersymmetry from Extra Dimensions,''
  Phys.\ Rev.\ D {\bf 94} (2016) no.9,  095017
  [arXiv:1608.06470 [hep-ph]].
  
\bibitem{Falk:1994es}
  T.~Falk, K.~A.~Olive and M.~Srednicki,
  ``Heavy sneutrinos as dark matter,''
  Phys.\ Lett.\ B {\bf 339} (1994) 248
  [hep-ph/9409270].

\bibitem{Arina:2007tm}
  C.~Arina and N.~Fornengo,
  ``Sneutrino cold dark matter, a new analysis: Relic abundance and detection rates,''
  JHEP {\bf 0711} (2007) 029
  [arXiv:0709.4477 [hep-ph]].


\bibitem{Tan:2016zwf}
  A.~Tan {\it et al.} [PandaX-II Collaboration],
  ``Dark Matter Results from First 98.7 Days of Data from the PandaX-II Experiment,''
  Phys.\ Rev.\ Lett.\  {\bf 117} (2016) no.12,  121303
  [arXiv:1607.07400 [hep-ex]].

\bibitem{Akerib:2016vxi}
  D.~S.~Akerib {\it et al.} [LUX Collaboration],
  ``Results from a search for dark matter in the complete LUX exposure,''
  Phys.\ Rev.\ Lett.\  {\bf 118} (2017) no.2,  021303
  [arXiv:1608.07648 [astro-ph.CO]].
  
\bibitem{Fox:2002bu}
  P.~J.~Fox, A.~E.~Nelson and N.~Weiner,
  ``Dirac gaugino masses and supersoft supersymmetry breaking,''
  JHEP {\bf 0208} (2002) 035
  [hep-ph/0206096].
  
\bibitem{Arina:2016rbb}
  C.~Arina, M.~Chala, V.~Martin-Lozano and G.~Nardini,
 ``Confronting SUSY models with LHC data via electroweakino production,''
  JHEP {\bf 1612} (2016) 149 [arXiv:1610.03822 [hep-ph]].

 
\bibitem{Han:2016xet}
  C.~Han, J.~Ren, L.~Wu, J.~M.~Yang and M.~Zhang,
  ``Top-squark in natural SUSY under current LHC run-2 data,''
  [arXiv:1609.02361 [hep-ph]].
  
  
  
\bibitem{Drees:1993uw}
  M.~Drees and M.~M.~Nojiri,
  ``Production and decay of scalar stoponium bound states,''
  Phys.\ Rev.\ D {\bf 49} (1994) 4595
  [hep-ph/9312213].

\bibitem{Martin:2008sv}
  S.~P.~Martin,
  ``Diphoton decays of stoponium at the Large Hadron Collider,''
  Phys.\ Rev.\ D {\bf 77} (2008) 075002
  [arXiv:0801.0237 [hep-ph]].

\bibitem{Bodwin:2016whr}
  G.~T.~Bodwin, H.~S.~Chung and C.~E.~M.~Wagner,
  ``Higgs-Stoponium Mixing Near the Stop-Antistop Threshold,''
  [arXiv:1609.04831 [hep-ph]].
 
  
  
\bibitem{Carena:2012gp}
  M.~Carena, S.~Gori, N.~R.~Shah, C.~E.~M.~Wagner and L.~T.~Wang,
  ``Light Stau Phenomenology and the Higgs $\gamma\gamma$ Rate,''
  JHEP {\bf 1207} (2012) 175
  [arXiv:1205.5842 [hep-ph]].
 
\bibitem{Marandella:2005wc}
  G.~Marandella, C.~Schappacher and A.~Strumia,
  ``Supersymmetry and precision data after LEP2,''
  Nucl.\ Phys.\ B {\bf 715} (2005) 173
  [hep-ph/0502095].

\bibitem{atl:2016sab} The ATLAS Collaboration, ``Search for the
  Supersymmetric Partner of the Top Quark in the Jets+Emiss Final
  State at sqrt(s) = 13 TeV,'' ATLAS-CONF-2016-077.

\bibitem{cms:2016srd} The CMS Collaboration, ``Search for direct top
  squark pair production in the fully hadronic final state in
  proton-proton collisions at sqrt(s) = 13 TeV corresponding to an
  integrated luminosity of 12.9/fb,'' CMS-PAS-SUS-16-029.

\bibitem{atl:2016src} The ATLAS Collaboration, ``Search for top-squark
  pair production in final states with two tau leptons, jets, and
  missing transverse momentum in $\sqrt{s}=13$ TeV pp-collisions with
  the ATLAS detector,'' ATLAS-CONF-2016-048.
  
   

\bibitem{Alves:2013wra} 
  D.~S.~M.~Alves, J.~Liu and N.~Weiner,
  ``Hiding Missing Energy in Missing Energy,''
  JHEP {\bf 1504}, 088 (2015)
  [arXiv:1312.4965 [hep-ph]].

\bibitem{Conte:2012fm} 
  E.~Conte, B.~Fuks and G.~Serret,
  ``MadAnalysis 5, A User-Friendly Framework for Collider Phenomenology,''
  Comput.\ Phys.\ Commun.\  {\bf 184}, 222 (2013)
  [arXiv:1206.1599 [hep-ph]].
  
\bibitem{Conte:2014zja} 
  E.~Conte, B.~Dumont, B.~Fuks and C.~Wymant,
  ``Designing and recasting LHC analyses with MadAnalysis 5,''
  Eur.\ Phys.\ J.\ C {\bf 74}, no. 10, 3103 (2014)
  [arXiv:1405.3982 [hep-ph]].
  
\bibitem{Brun:1997pa} 
  R.~Brun and F.~Rademakers,
  ``ROOT: An object oriented data analysis framework,''
  Nucl.\ Instrum.\ Meth.\ A {\bf 389}, 81 (1997).

\bibitem{Cacciari:2011ma} 
  M.~Cacciari, G.~P.~Salam and G.~Soyez,
  ``FastJet User Manual,''
  Eur.\ Phys.\ J.\ C {\bf 72}, 1896 (2012)
  [arXiv:1111.6097 [hep-ph]].
  
\bibitem{Read:2002hq} 
  A.~L.~Read,
  ``Presentation of search results: The CL(s) technique,''
  J.\ Phys.\ G {\bf 28}, 2693 (2002).


\bibitem{Alwall:2014hca} 
  J.~Alwall {\it et al.},
  ``The automated computation of tree-level and next-to-leading order differential cross sections, and their matching to parton shower simulations,''
  JHEP {\bf 1407}, 079 (2014)
  [arXiv:1405.0301 [hep-ph]].
  
\bibitem{Sjostrand:2006za} 
  T.~Sjostrand, S.~Mrenna and P.~Z.~Skands,
  ``PYTHIA 6.4 Physics and Manual,''
  JHEP {\bf 0605}, 026 (2006)
  [hep-ph/0603175].


 
\bibitem{Staub:2013tta}
  F.~Staub,
  ``SARAH 4 : A tool for (not only SUSY) model builders,''
  Comput.\ Phys.\ Commun.\  {\bf 185} (2014) 1773
  [arXiv:1309.7223 [hep-ph]].

\bibitem{Porod:2011nf}
  W.~Porod and F.~Staub,
  ``SPheno 3.1: Extensions including flavour, CP-phases and models beyond the MSSM,''
  Comput.\ Phys.\ Commun.\  {\bf 183} (2012) 2458
  [arXiv:1104.1573 [hep-ph]].


\bibitem{Capela:2012uk}
  F.~Capela and G.~Nardini,
  ``Hairy Black Holes in Massive Gravity: Thermodynamics and Phase Structure,''
  Phys.\ Rev.\ D {\bf 86} (2012) 024030
    [arXiv:1203.4222 [gr-qc]].


\bibitem{Clesse:2016vqa}
  S.~Clesse and J.~Garcia-Bellido,
  ``The clustering of massive Primordial Black Holes as Dark Matter: measuring their mass distribution with Advanced LIGO,''
  Phys.\ Dark Univ.\  {\bf 10} (2016) 002
    [arXiv:1603.05234 [astro-ph.CO]].

\bibitem{Bird:2016dcv}
  S.~Bird, I.~Cholis, J.~B.~Munoz, Y.~Ali-Hamoud, M.~Kamionkowski, E.~D.~Kovetz, A.~Raccanelli and A.~G.~Riess,
  ``Did LIGO detect dark matter?,''
  Phys.\ Rev.\ Lett.\  {\bf 116} (2016) no.20,  201301
   [arXiv:1603.00464 [astro-ph.CO]].

\bibitem{McGaugh:2016leg}
  S.~McGaugh, F.~Lelli and J.~Schombert,
  ``Radial Acceleration Relation in Rotationally Supported Galaxies,''
  Phys.\ Rev.\ Lett.\  {\bf 117} (2016) no.20,  201101
  [arXiv:1609.05917 [astro-ph.GA]].

\bibitem{Gelmini:2010zh}
  G.~Gelmini and P.~Gondolo,
  ``DM Production Mechanisms,''
  In *Bertone, G. (ed.): Particle dark matter* 121-141
  [arXiv:1009.3690 [astro-ph.CO]].

\bibitem{Nardini:2011hu}
  G.~Nardini and N.~Sahu,
  ``Re-Reheating, Late Entropy Injection and Constraints from Baryogenesis Scenarios,''
  [arXiv:1109.2829 [hep-ph]].

\bibitem{Dreiner:1997uz}
  H.~K.~Dreiner,
  ``An Introduction to explicit R-parity violation,''
  Adv.\ Ser.\ Direct.\ High Energy Phys.\  {\bf 21} (2010) 565
  [hep-ph/9707435].



\end{thebibliography}
\end{document}